%% file: held-karp.tex
\documentclass[11pt,leqno]{article}
\usepackage[letterpaper]{geometry} %
\usepackage{article} %
\usepackage{math} %
\usepackage{algo} %
\usepackage{wrapfig} %
\usepackage{graphicx} %
\usepackage{placeins} %
\usepackage{multicol} %
\usepackage{titlesec} %
\usepackage[font=small,labelfont=it]{caption}

\titleformat{\subparagraph}[runin]{\itshape}{}{0pt}{\noindent}{}

\titleformat{\paragraph}[runin]{\bfseries}{}{0pt}{\noindent}{}

\renewcommand{\subparagraph}[1]{\medskip\noindent\textit{#1}}
\renewcommand{\paragraph}[1]{\medskip \noindent {\bf #1}} %

\newcommand{\epsmore}{\parof{1 + \eps}} %
\newcommand{\cuts}{\mathcal{C}} %
\newcommand{\lca}{\operatorname{\algo{lca}}\optpar} %
\newcommand{\incomp}{\parallel} %

\newcommand{\packing}{p} \newcommand{\cut}{\mathscr{C}\optpar} %
\newcommand{\cutw}{\overline{\mathscr{C}}\optpar} %
\newcommand{\cutweight}{\cutw} %
\newcommand{\downset}{D\optpar} %
\newcommand{\mincap}{\gamma} %

\newcommand{\canonicalcuts}[1][T]{\mathcal{D}_{#1}} %

\newcommand{\mtsp}{{\sf Metric-TSP}\xspace} %
\newcommand{\tsp}{{\sf TSP}\xspace} %
\newcommand{\gtsp}{{\sf Graphic-TSP}\xspace} %
\newcommand{\ecss}{{\sf 2ECSS}\xspace} %
\newcommand{\ecssd}{{\sf 2ECSSD}\xspace} %
\newcommand{\atsp}{{\sf Asymmetric-TSP}\xspace} %
\newcommand{\tsppath}{{\sf TSP-Path}\xspace} %

\renewcommand{\cite}{\citeyearpar}

\begin{document}

\title{Approximating the \mbox{Held-Karp Bound} \\ for \mbox{Metric
    TSP} \mbox{in Nearly Linear Time}\thanks{Department of Computer
    Science, University of Illinois, Urbana, IL
    61820. \{chekuri,quanrud2\}@illinois.edu.  Work on this paper
    partially supported by NSF grant CCF-1526799.}  }

\author{Chandra Chekuri \and Kent Quanrud}

\pagenumbering{gobble}

\maketitle

\begin{abstract}
  We give a nearly linear time randomized approximation scheme for the
  Held-Karp bound \citep*{hk-70} for \mtsp. Formally, given an
  undirected edge-weighted graph $\defgraph$ on $m$ edges and
  $\eps > 0$, the algorithm outputs in $O(m \log^4n /\eps^2)$ time,
  with high probability, a $(1+\eps)$-approximation to the Held-Karp
  bound on the \mtsp instance induced by the shortest path metric on
  $\graph$. The algorithm can also be used to output a corresponding
  solution to the Subtour Elimination LP. We substantially improve
  upon the $O(m^2 \log^2(m)/\eps^2)$ running time achieved previously
  by Garg and Khandekar. The LP solution can be used to obtain a fast
  randomized $\parof{\frac{3}{2} + \eps}$-approximation for \mtsp
  which improves upon the running time of previous implementations of
  Christofides' algorithm.
  % Along the way we obtain a data structure for
  % incremental maintenance of a global mincut of a weighted graph
  % with
  % poly-logarithmic amortized query time.
  % The algorithm builds on our recent framework for accelerating
  % multiplicative weight update based methods for implicit fractional
  % packing problems.
\end{abstract}

\newpage

\pagenumbering{arabic}

\section{Introduction}
\labelsection{intro}

The \emph{Traveling Salesman Problem} (\tsp) is a central problem in
discrete and combinatorial optimization, and has inspired fundamental
advances in optimization, mathematical programming and theoretical
computer science. Cook's recent book \cite{c-14} gives an introduction
to the problem, its history, and general appeal. See also
\citet{gp-06}, \citet{abcc-tspbook-11}, and \citet{llrs-85} for
book-length treatments of \tsp and its variants.

Formally, the input to \tsp is a graph $\defgraph$ equipped with
positive edge costs $\capacity: \edges \to \preals$. The goal is to
find a minimum cost Hamiltonian cycle in $\graph$. In this paper we
focus on \tsp in undirected graphs. Checking whether a given graph has
a Hamiltonian cycle is a classical NP-Complete decision problem, and
hence \tsp is not only NP-Hard but also inapproximable. For this
theoretical reason, as well as many practical applications, a special
case of \tsp called \mtsp is extensively studied.  In \mtsp, $\graph$
is a complete graph $K_n$ and $\capacity$ obeys the triangle
inequality $\capacity{uv} \leq \capacity{uw} + \capacity{wv}$ for all
$u,v,w \in \vertices$. An alternative interpretation of \mtsp is to
find a minimum-cost \emph{tour} of an edge-weighted graph $\graph$;
where a tour is a closed walk that visits all the vertices.  In other
words, \mtsp is a relaxation of \tsp in which a vertex can be visited
more than once. The graph-based view of \mtsp allows one to specify
the metric on $\vertices$ implicitly and \emph{sparsely}.

Unlike \tsp, which is inapproximable, \mtsp admits a constant factor
approximation. The classical algorithm of \citet{c-76} yields a
$3/2$-approximation. On the other hand it is known that \mtsp is
APX-Hard and hence does not admit a PTAS (\citet{l-12} showed that
there is no $\frac{185}{184}$-approximation unless $P=NP$). An
outstanding open problem is to improve the bound of $3/2$.  A
well-known conjecture states that the worst-case integrality gap of
the \emph{Subtour-Elimination LP} formulated by \citet*{dfj-54} is
$4/3$ (see \citep{g-tsp-95}).  There has been exciting recent progress
on this conjecture and several related problems; we refer the reader
to an excellent survey by \citet{v-12}.  The Subtour Elimination LP
for \tsp is described below and models the choice to take an edge
$e \in \edges$ with a variable $y_e \in [0,1]$. In the following, let
$\cut{U}$ (resp.\ $\cut{v}$) denotes the set of edges crossing the set
of vertices $U \subseteq \vertices$ (resp.\ the vertex
$v \in \vertices$).
\begin{align*}
  \begin{array}{r l l}
    \operatorname{SE}(\graph,\capacity) = \min %
    & \ripof{\capacity}{y}        %
    & \\
    \text{s.t.}%
    &                             %
      \sum_{e \in \cut{v}} y_e = 2
    &
      \text{for all } v \in \vertices,\\
    & \sum_{e \in \cut{U}} y_e \geq 2
    & \text{for all } %
      \emptyset \subsetneq U \subsetneq \vertices,\\
    \text{and} %
    & y_e \in [0,1] & \text{for all } e \in \edges.
  \end{array}
\end{align*}
The first set of constraints require each vertex to be incident to
exactly two edges (in the integral setting); these are referred to as
degree constraints. The second set of constraints force connectivity,
hence the name ``subtour elimination''.  The LP provides a lower bound
for \tsp, and in order to apply it to an instance of \mtsp defined by
$\graph$, one needs to apply it to the metric completion of $\graph$.

A problem closely related to \mtsp is the \emph{$2$-edge-connected
  spanning subgraph problem} (\ecss). In \ecss the input is an
edge-weighted graph $(\defgraph, \capacity)$, and the goal is to find
a minimum cost subgraph of $\graph$ that is 2-edge-connected. We focus
on the simpler version where an edge is allowed to be used more than
once. A natural LP relaxation for \ecss is described below on the
left. We have a variable $y_e$ for each edge $e \in \edges$, and
constraints which ensure that each cut has at least two edges crossing
it. We also describe the dual LP on the right which corresponds to a
maximum packing of cuts into the edge costs. In the following, let
$\cuts \subseteq \subsetsof{\edges}$ denote the family of all cuts in
$\graph$. (For technical reasons, we prefer to treat cuts as sets of
edges.)

\bigskip

\hfill%
\begin{minipage}{.4\textwidth}
  \begin{framed}
    \vspace{-1em}
    \begin{align*}
      & \hspace{-8ex} \ecss(\graph,\capacity) = \min
        \ripof{\capacity}{y}  \hspace{-\textwidth} \\
      \operatorname{s.t.\ } & \sum_{e \in C} y_e \geq 2 & \forall C
                                                          \in \cuts \\
      \text{and\ } & y_e \geq 0 & \forall e \in \edges
    \end{align*}
    \vspace{-3.5ex}
  \end{framed}
\end{minipage}%
\hfill%
\begin{minipage}{.4\textwidth}
  \begin{framed}
    \vspace{-1em}
    \begin{align*}
      & \hspace{-8ex} \ecssd(\graph,\capacity) =
        \text{max } 2 \ripof{\ones}{x} \hspace{-\textwidth} \\
      \operatorname{s.t.\ } & \sum_{C \ni e} x_C \leq \capacity{e}
                            & \forall e \in \edges \\
      \text{and\ } & x_C \geq 0 & \forall C \in \cuts
    \end{align*}                %
    \vspace{-3.5ex} %
  \end{framed}
\end{minipage}%
\hfill\hfill

\bigskip

Cunningham (see \citep{mmp-90}) and \citet{gb-93} observed that for
any edge-weighted graph $(\graph,\capacity)$, the optimum value of the
Subtour Elimination LP for the metric completion of
$(\graph,\capacity)$ coincides with the optimum value of the $\ecss$
LP for $(\graph,\capacity)$.  The advantage of this connection is
twofold. First, the \ecss relaxation is a pure covering LP, and its
dual is a pure packing LP. Second, the \ecss formulation works
directly with the underlying graph $(\graph,\capacity)$ instead of the
metric completion.

\medskip
\noindent {\bf On the importance of solving the Subtour-LP:} The
subtour elimination LP is extensively studied in mathematical
programming both for its application to TSP as well as the many
techniques its study has spawned. It is a canonical example in many
books and courses on linear and integer programming. The seminal paper
of Dantzig, Fulkerson and Johnson proposed the cutting plane method
based on this LP as a way to solve TSP exactly. \citet{abcc-03}
demonstrated the power of this methodology by solving TSP on extremely
large real world instances; the resulting code named Concorde is
well-known \citep{abcc-tspbook-11}.  The importance of solving the
subtour elimination LP to optimality has been recognized since the
early days of computing. The Ellipsoid method can be used to solve the
LP in polynomial time since the separation oracle required is the
global mincut problem. However, it is not practical.  One can also
write polynomial-sized extended formulations using flow variables, but
the number of variables and constraints is cubic in $n$ and this too
leads to an impractical algorithm.  \citet{hk-70} provided an
alternative lower bound for \tsp via the notion of
\emph{one-trees}. They showed, via Lagrangian duality, that their
lower bound coincides with the one given by
$\operatorname{SE}\parof{\graph,\capacity}$. The advantage of the
Held-Karp bound is that it can be computed via a simple iterative
procedure relying on minimum spanning tree computations.  In practice,
this iterative procedure provides good estimates for the lower
bound. However, there is no known polynomial-time implementation with
guarantees on the convergence rate to the optimal value.

In the rest of the paper we focus on \mtsp. For the sake of brevity,
we refer to the Held-Karp bound for the metric completion of
$(\graph,\capacity)$ as simply the Held-Karp bound for
$(\graph,\capacity)$. How fast can one compute the Held-Karp bound for
a given instance? Is there a strongly polynomial-time or a
combinatorial algorithm for this problem? These questions have been
raised implicitly and are also explicitly pointed out, for instance,
in \citep{bp-90} and \citep{gb-93}.  A fast algorithm has several
applications ranging from approximation algorithms to exact algorithms
for TSP.

\citet*{pst-95}, in their influential paper on fast approximation
schemes for packing and covering LPs via Lagrangian relaxation
methods, showed that a $(1+\eps)$-approximation for the Held-Karp
bound for \mtsp can be computed in $\bigO{n^4\log^6 n/\eps^2}$
randomized time. They relied on an algorithm for computing the global
minimum cut\footnote{Their scheme can in fact be implemented in
  randomized $O(m^2 \log^4m/\eps^2)$ time using subsequent
  developments in minimum cut algorithms and width reduction
  techniques.}. Subsequently, Garg and Khandekar obtained a
$(1+\eps)$-approximation in $O(m^2 \log^2 m/\eps^2)$ time and they
relied on algorithms for minimum-cost branchings (see
\citep{khandekar-04}).

\medskip
\noindent {\bf The main result.} In this paper we obtain a near-linear
running time for a $(1+\eps)$-approximation, substantially improving
the best previously known running time bound.

\begin{theorem}
  \labeltheorem{main} Let $\defgraph$ be an undirected graph with
  $\sizeof{\edges} = m$ edges and $\sizeof{\vertices} = n$ vertices,
  and positive edge weights $\capacity: \edges \to \preals$. For any
  fixed $\eps > 0$, there exists a randomized algorithm that computes
  a $(1 + \eps)$-approximation to the Held-Karp lower bound for the
  \mtsp instance on $(\graph, \capacity)$ in $O(m \log^4 n/ \eps^2)$
  time. The algorithm succeeds with high probability.
\end{theorem}

The algorithm in the preceding theorem can be modified to return a
$(1+\eps)$-approximate solution to the \ecss LP within the same
asymptotic time bound.  For fixed $\eps$, the running time we achieve
is asymptotically faster than the time to compute or even write down
the metric completion of $(\graph,\capacity)$. Our algorithm can be
applied low-dimensional geometric point sets to obtain a running-time
that is near-linearly in the number of points.

In typical approximation algorithms that rely on mathematical
programming relaxations, the bottleneck for the running time is
solving the relaxation.  Surprisingly, for algorithms solving \mtsp
via the Held-Karp bound, the bottleneck is no longer solving the
relaxation (albeit we only find a $(1+\eps)$-approximation and do not
guarantee a basic feasible solution). We mention that the recent
approaches towards the $4/3$ conjecture for \mtsp are based on
variations of the classical Christofides heuristic (see
\citep{v-12}). The starting point is a near-optimal feasible solution
$x$ to the \ecss LP on $(\graph,\capacity)$.  Using a well-known fact
that a scaled version of $x$ lies in the spanning tree polytope of
$\graph$, one generates one or more (random) spanning trees $T$ of
$\graph$.  The tree $T$ is then augmented to a tour via a min-cost
matching $M$ on its odd degree nodes. \citet{gw-17} recently evaluated
some of these {\em Best-of-Many Christofides' algorithms} and
demonstrated their effectiveness.  A key step in this scheme, apart
from solving the LP, is to decompose a given point $y$ in the spanning
tree polytope of $\graph$ into a convex combination of spanning
trees. Our recent work \citep{cq-17} shows how to achieve a
$(1-\eps)$-approximation for this task in near-linear time; the
algorithm implicitly stores the decomposition in near-linear
space. One remaining bottleneck to achieve an overall near-linear
running time is to compute an approximate min-cost perfect matching on
the odd-degree nodes of a given spanning tree $T$.  In recent work
\citep{ChekuriQ17b}, we have been able to overcome this bottleneck in
one way.  We obtain a randomized algorithm which uses a feasible
solution $x$ to \ecss LP as input, and outputs a perfect matching $M$
on the odd-degree nodes of $T$ whose expected cost is at most
$(\frac12 + \eps)$ times the cost of $x$.  Combined with our algorithm
in \reftheorem{main}, this leads to a
$\parof{\frac{3}{2} + \eps}$-approximation for \mtsp in
$\apxO(m/\eps^2 + n^{1.5}/\eps^3)$ time. If the metric space is given
explicitly, then the overall run time is $\apxO(n^2/\eps^2)$ and
near-linear in the input size. Previous implementations of
Christofides' algorithm required $\Omega(n^{2.5}\log \frac{1}{\eps})$
time to obtain a $(\frac{3}{2} + \eps)$-approximation even when the
metric space is given explicitly.

\iffalse

{\bf Skipped the discussion on geometric instances.  Appears to
  distract and undercut the graph result in some ways.}

We remark that our algorithm can also be used for geometric instances.
Given a \mtsp instance induced by a set $P$ of $n$ points in
$\mathbb{R}^d$ equipped with the Euclidean metric, and an $\eps > 0$,
one can compute in $\bigO{n \log n + n \eps^{-d}}$ time, a graph on
$P$ that is a $(1+\eps)$-approximate spanner, and has
$\bigO{n \eps^{-d}}$ edges~\citep[Theorem 3.12]{h-11}; this also
extends to points with doubling dimension $d$. Thus, the Held-Karp
bound for $P$ can be computed in $\apxO{n\eps^{-(d+2)}}$ time. It may
be feasible to improve the dependence on $\eps$ since we do not need
the full strength of a $(1+\eps)$-spanner. A $(1+\eps)$-approximation
for \mtsp (in terms of the actual integral optimum) in Euclidean
spaces can be computed in $O(2^{(d/\eps)^{O(d)}}n)$ time
\citep{bg-tsp-13,rs-tsp-98}. The exponential dependence on $1/\eps$ is
necessary (unless $P = NP$).  \fi

\subsection{Integrated design of the algorithm}

\iffalse Implicit in the proof of \reftheorem{main} is a data
structure that maintains a $(1+\eps)$-approximate global mincut of a
partially dynamic graph where edge-weights only increase. We achieve
amortized polylogarithmic time for edge weight updates and maintain a
compact representation of an approximately minimum min-cut. The actual
edges of the approximate minimum cut can be reported in
polylogarithmic time plus constant time per edge reported. This is not
the main focus of the paper and we do not explicitly state or work out
a formal interface to this data structure.  \fi

% {\bf Can we solve the s-t path TSP LP as well? What about rounding?}

% \paragraph{Integrated design of the algorithm.}

Our algorithm is based on the multiplicative weight update framework
(MWU), and like \citet*{pst-95}, we approximate the pure packing LP
\ecssd.  Each iteration requires an oracle for computing the global
minimum cut in an undirected graph.  A single minimum cut computation
takes randomized near-linear-time via the algorithm of \citep{k-00},
and the MWU framework requires $\apxOmega(m/\eps^2)$ iterations.
Suprisingly, the whole algorithm can be implemented to run in roughly
the same time as that required to compute one global mincut.
% We employ two key high-level ideas.

While the full algorithm is fairly involved, the high-level design is
directed by some ideas developed in recent work by the authors
\citep{cq-17} that is inspired by earlier work of \citet{m-10} and
\citet{y-14}. We accelerate MWU-based algorithms for some implicit
packing problems with the careful interplay of two data
structures. The first data structure maintains a minimum cost object
of interest (here the global minimum cut) by (partially) dynamic
techniques, rather than recompute the object from scratch in every
iteration. The second data structure applies the multiplicative weight
update in a lazy fashion that can be amortized efficiently against the
weights of the constraints. The two data structures need to be
appropriately meshed to obtain faster running times, and this meshing
depends very much on the problem specifics as well as the details of
the dynamic data structures.

While we do inherit some basic ideas and techniques from this
framework, the problem here is more sophisticated than those
considered in \citep{cq-17}. The first component in this paper is a
fast dynamic data structure for maintaining a $(1+\eps)$-approximate
global minimum cut of a weighted graph whose edge weights are only
increasing. We achieve an amortized poly-logarithmic update time by a
careful adaptation of the randomized near-linear-time minimum cut
algorithm of \citet{k-00} that relies on approximate tree packings.
This data structure is developed with careful consideration of the MWU
framework; for example, we only need to compute an approximate tree
packing $\apxO{1 / \eps^2}$ times (rather than in every iteration)
because of the monotonicity of the weight updates and standard upper
bounds on the total growth of the edge weights.

The second technical ingredient is a data structure for applying a
multiplicative weight update to each edge in the approximately minimum
cuts selected in each iteration. The basic difficulty here is that we
cannot afford to touch each edge in the cut. While this task suggests
a lazy weight update similar to \citep{cq-17}, these techniques
require a compact representation of the minimum cuts. It appears
difficult to develop in isolation a data structure that can apply
multiplicative weight updates along any (approximately) minimum cut.
However, additional nice properties of the cuts generated by the first
dynamic data structure enable a clean interaction with lazy weight
updates. We develop an efficient data structure for weight updates
that is fundamentally inextricable from the data structure generating
the approximately minimum cuts.

\begin{remark}
  If we extract the data structure for minimum cuts from the MWU
  framework, then we obtain the following.
  \begin{quote}
    \itshape Given an edge-weighted $G$ graph on $m$ edges there is a
    dynamic data structure for the weighted incremental maintenance of
    a $(1+\eps)$-approximate global minimum cut that updates and
    queries in constant time plus
    $O(m \polylog{n}\logof{W_1 / W_0}/\eps)$ total amortized time,
    where $W_0$ is the weight of the minimum cut of the initial graph
    and $W_1$ is the weight of the minimum cut of the final graph
    (after all updates).  Here, updates in the weighted incremental
    setting consist of an edge $e$ and a positive increment $\alpha$
    to its weight.  The edges of the approximate minimum cut can be
    reported in constant time per edge reported.
  \end{quote}
  However, a data structure for dynamic maintenance of minimum cuts is
  not sufficient on its own, as there are several subtleties to be
  handled both appropriately and efficiently. For instance,
  \citet{t-07} developed a {\em randomized fully-dynamic} data
  structure for global mincut with a worst-case update time of
  $\tilde{O}(\sqrt{n})$. Besides the slower update time, this data
  structure assumes a model and use case that clashes with the MWU
  framework at two basic points. First, the data structure does not
  provide access to the edges of the mincut in an implicit fashion
  that allows the edge weights to be updated efficiently, and updating
  each of the edges of the minimum cut individually leads to quadratic
  running times. Second, the randomization in the data structure
  assumes an oblivious adversary, which is not suitable for our
  purposes where the queries of the MWU algorithm to the data
  structure are {\em adaptive}.
\end{remark}

There are many technical details that the above overview necessarily
skips.  We develop the overall algorithm in a methodical and modular
fashion in the rest of the paper.

\subsection{Related Work}
Our main result has connections to several important and recent
directions in algorithms research. We point out the high-level aspects
and leave a more detailed discussion to a later version of the paper.
A reader more interested in the algorithm should feel free to skip
this subsection and directly go to the next section.

In one direction there has been a surge of interest in faster
(approximate) algorithms for classical problems in combinatorial
optimization including flows, cuts, matchings and linear programming
to name a few. These are motivated by not only theoretical
considerations and techniques but also practical applications with
very large data sizes. Many of these are based on the interplay
between discrete and continuous techniques and there have been several
breakthroughs, too many to list here. One prominent recent example of
a problem that admits a near-linear-time approximation scheme is
maximum flow in undirected graphs \citep{p-16} building on many
previous tools and techniques. Our work adds to the list of basic
problems that admit a near-linear-time approximation scheme.

In another direction there has been exciting work on approximation
algorithms for \tsp and its variants including \mtsp, \gtsp, \atsp,
\tsppath to name a few. Several new algorithms, techniques,
connections and problems have arisen out of this work. Instead of
recapping the many results we refer the reader to the survey
\citep{v-12}.  Our result adds to this literature and suggests that
one can obtain not only improved approximation but also much faster
approximations than what was believed possible.  As we already
mentioned, some other ingredients in the rounding of a solution such
as decomposing a fractional point in a spanning tree polytope into
convex combination of spanning trees can also be sped up using some of
the ideas in our prior work.

In terms of techniques, our work falls within the broad framework of
improved implementations of iterative algorithms.  Our recent work
showed that MWU based algorithms, especially for implicit packing and
covering problems, have the potential to be substantially improved by
taking advantage of data structures. The key, as we described already,
is to combine multiple data structures together while exploiting the
flexibility of the MWU framework. Although not a particularly novel
idea (see the work of \citet{m-10} for applications to multicommodity
flow and the work of \citet{ap-14} for geometric covering problems),
our work demonstrates concrete and interesting problems to highlight
the sufficient conditions under which this is
possible. Lagrangian-relaxation based approximation schemes for
solving special classes of linear programs have been studied for
several decades with a systematic investigation started in
\citet*{pst-95} and \citet{gk-94} following earlier work on
multicommodity flows. Since then there have been many ideas,
refinements and applications. We refer the reader to
\citep{ky-efpc-14,y-14,cjv-15} for some recent papers on MWU-based
algorithms which supply further pointers, and to the survey
\citep{ahk-12} for the broader applicability of MWU in theoretical
computer science. Accelerated gradient descent methods have recently
resulted in fast randomized near-linear-time algorithms for explicit
fractional packing and covering LPs; these algorithms improved the
dependence of the running time on $\eps$ to $1/\eps$ from
$1/\eps^2$. See \citep{ao-plps-15} and \citep{wrm-uam-15}.

Our work here builds extensively on Karger's randomized nearly linear
time mincut algorithm \cite{k-00}. As mentioned already, we adapt his
algorithm to a partially dynamic setting informed by the MWU
framework. \citep{t-07} also builds on Karger's tree packing ideas to
develop a \emph{fully} dynamic mincut algorithm.  Thorup's algorithm
is rather involved and is slower than what we are able to achieve for
the partial dynamic setting; he achieves an update time of
$\apx{O}(\sqrt{n})$ while we achieve polylogarithmic time.  There are
other obstacles to integrating his ideas and data structure to the
needs of the MWU framework as we already remarked. For unweighted
incremental mincut, \citet{ght-16} recently developed a deterministic
data structure with a poly-logarithmic amortized update time.

Some of the data structures in this paper rely on Euler tour
representations of spanning trees. The Euler tour representation
imposes a particular linear order on the vertices by which the edges
of the underlying graph can be interpreted as intervals. The Euler
tour representation was introduced by \citet{tv-85} and has seen
applications in other dynamic data structures, such as the work by
\citet{hk-99} among others.

\paragraph{Organization.}
Our main result is a combination of some high-level ideas and several
data structures. The implementation details take up considerable
space. Instead of compressing the details we organized the paper in a
modular fashion with the understanding that the reader may skip some
details here and there.  \refsection{mwu} gives a high-level overview
of a first MWU-based algorithm, highlighting some important properties
of the MWU framework that arise in more sophisticated arguments later.
\refsection{tree-packings} summarizes the key properties of tree
packings that underlie Karger's mincut algorithm, and explains their
use in our MWU algorithm via the notion of epochs.
\refsection{tree-cuts} describes an efficient subroutine, via
appropriate data structures, to find all approximate mincuts (in the
partial dynamic setting) induced by the spanning trees of a tree
packing. \refsection{lazy-cut-weights} describes the data structure
that implement the weights of the edges in a lazy fashion, building on
Euler tours of spanning trees, range trees, and some of our prior
work. In \refsection{putting-together}, we outline a formal proof of
\reftheorem{main}.

\section{MWU based Algorithm and Overview}
\labelsection{mwu}

\begin{figure}[t]
  \input{algos/held-karp-direct}
  \caption{An exact (and slow) implementation of the MWU framework
    applied to packing cuts.  \labelfigure{held-karp-direct}}

  \hrulefill
\end{figure}

Given an edge-weighted graph $(\defgraph,\capacity)$ our goal is to
find a $(1-\eps)$-approximation to the optimal value of the LP
\ecssd. We later describe how to obtain an approximate solution to the
primal LP \ecss. Note that the LP is a packing LP of the form
$\max \ripof{\ones}{x} \where Ax \le \capacity$ with an exponential
number of variables corresponding to the cuts, but only $m$
non-trivial constraints corresponding to the edges. The MWU framework
can be used to obtain a $(1-\eps)$-approximation for fractional
packing.  Here we pack cuts into the edge capacities $\capacity$.  The
MWU framework reduces packing cuts to finding global minimum cuts as
follows. The framework starts with an empty solution
$x = \zeroes^{\cuts}$ and maintains edge weights
$\weight: \edges \to \nnreals$, initialized to $1 / \capacity$ and
\emph{non-decreasing} over the course of the algorithm. Each
iteration, the framework finds the minimum cut
\begin{math}
  C = \argmin_{C' \in \cuts}\sumweight{C'}
\end{math}
w/r/t $\weight$, where
\begin{math}
  \sumweight{C'} \defeq \sum_{e \in C'} \weight{e}
\end{math}
denotes the total weight of a cut. The framework adds a fraction of
$C$ to the solution $x$, and updates the weight of every edge in the
cut in a multiplicative fashion such that the weight of an edge is
exponential in its load $x(e)/c(e)$. The weight updates steer the
algorithm away from reusing highly loaded edges.  The MWU framework
guarantees that the output $x$ will have objective value
$\ripof{\ones}{x} \geq \opt$ while satisfying
\begin{math}
  \sum_{C \ni e} x_C \leq (1 + O(\eps))\capacity{e}
\end{math}
for every edge $e$. Scaling down $x$ by a $(1 + O(\eps))$-factor gives
a $(1 - O(\eps))$-approximation satisfying all packing constraints.

An actual implementation of the above high-level idea needs to specify
the step size in each iteration and the precise weight-update. The
non-uniform increments idea of \citet{gk-07} gives rise to a
width-independent bound on the number of iterations, namely
$O(m \log m/\eps^2)$ in our setting. Our algorithms follow the
specific notation and scheme of \citet{cjv-15}, which tracks the
algorithm's progress by a ``time'' variable $t$ from $0$ to $1$
increasing in non-uniform steps. A step of size $\delta$ in an
iteration corresponds to adding $\delta \beta 1_C$ to the current
solution $x$ where $1_C$ is the characteristic vector of the mincut
$C$ found in the iteration, and
$\beta = \ripof{\weight}{\capacity} / \sumweight{C}$ greedily takes as
much of the cut as can fit in the budget $\ripof{\weight}{\capacity}$.
The process is controlled by a parameter $\eta$, which when set to
$O(\ln m/\eps)$ guarantees both the width-independent running time
bound as well as the $(1-\eps)$-approximation.

Both the correctness and the number of iterations are derived by a
careful analysis of the edge weights $\weight$. We review the argument
that bounds the number of iterations. At the beginning of the
algorithm, when every edge $e$ has weight
$\weight{e} = 1 / \capacity{e}$, we have
$\ripof{\weight}{\capacity} = m$. The weights monotonically increase,
and standard proofs show that this inner product is bounded above by
\begin{math}
  \ripof{\weight}{\capacity} \leq \apxO{m^{O(1/\eps)}}.
\end{math}
The edge weight increases are carefully calibrated so that at least
one edge increases by a $(1 + \eps)$-multiplicative factor in each
iteration. As the upper bound on $\ripof{\weight}{\capacity}$ is an
upper bound on $\weight{e}\capacity{e}$ for any edge $e$, and the
initial weight of an edge $e$ is $1/\capacity{e}$, an edge weight
$\weight{e}$ can increase by a $(1+\eps)$-factor at most
\begin{math}
  \apxlog{m^{O(1/\eps)}} %
  = %
  \bigO{\logof{m}/\eps^2}
\end{math}
times. Charging each iteration to an edge weight increased by a
$(1+\eps)$-factor, the algorithm terminates in $\apxO{m/\eps^2}$
iterations.

A direct implementation of the MWU framework,
\refroutine{held-karp-1}, is given in \reffigure{held-karp-direct}. In
addition to the input $(\graph, \capacity, \eps)$, there is an
additional parameter $\eta > 0$ that dampens the step size at each
iteration. Standard analysis shows that the appropriate choice of
$\eta$ is about $\lnof{m} / \eps$. Each iteration requires the minimum
global cut w/r/t the edge weights $\weight$, which can be found in
$\apxO{m}$ time (with high probability) by the randomized algorithm of
\citet{k-00}. Calling Karger's algorithm in each iteration gives a
quadratic running time of $\apxO{m^2 / \eps^2}$.

\paragraph{Approximation in the framework.} The MWU framework is
robust to approximation in several ways, especially in the setting of
packing and covering where non-negativity is quite helpful. For
example, it suffices to find a $(1 + O(\eps))$-multiplicative
approximation of the minimum cut, maintain the edge weights to a
$(1 + O(\eps))$-multiplicative approximation of their true edge
weights, and so forth, while retaining a
$(1 + O(\eps))$-multiplicative approximation overall. We exploit this
slack in two concrete ways that we describe at a high-level below.

\begin{figure}[t]
  \input{algos/held-karp-epochs}
  \caption{A second implementation of the MWU framework applied to
    packing cuts that divides the iterations into epochs and seeks
    only approximately minimum cuts at each iteration.}

  \hrulefill
\end{figure}

\subparagraph{Approximate mincuts.} First, we only look for cuts that
are within a $(1+O(\eps))$-approximate factor of the true minimum
cut. To this end, we maintain a target cut value $\lambda > 0$, and
maintain the invariant that there are no cuts of value strictly less
than $\lambda$. We then look for cuts of value
$\leq (1 + O(\eps))\lambda$ until we are sure that there are no more
cuts of value $\leq (1 + \eps)\lambda$. When we have certified that
there are no cuts of value $\leq (1+\eps) \lambda$, we increase
$\lambda$ to $(1+\eps)\lambda$, and repeat. Each stretch of iterations
with the same target value $\lambda$ is called an \emph{epoch}.
Epochs were used by \citep{f-00} for approximating fractional
multicommodity flow. We show that \emph{all} the approximately minimum
cuts in a single epoch can be processed in $\apxO{m}$ total time (plus
some amortized work bounded by other techniques).  If $\kappa$ is the
weight of the initial minimum cut (when $\weight = 1/ \capacity$),
then the minimum cut size increases monotonically from $\kappa$ to
$\apxO{m^{1/\eps} \kappa}$. If the first target cut value is
$\lambda = \kappa$, then there are only $O(\ln m/\eps^2)$ epochs over
the course of the entire algorithm.

\subparagraph{Lazy weight updates.} Even if we can concisely identify
the approximate minimum cuts quickly, there is still the matter of
updating the weights of all the edges in the cut quickly.  A cut can
have $\Omega(m)$ edges, and updating $m$ edge weights in each of
$\apxO{m/\eps^2}$ iterations leads to a $\apxO{m^2/\eps^2}$ running
time.  Instead of visiting each edge of the cut, we have a subroutine
that simulates the weight increase to all the edges in the cut, but
only does work for each edge whose (true) weight increases by a full
$(1+\eps)$-multiplicative power. That is, for each edge $e$, we do
work proportional (up to log factors) to the number of times
$\weight{e}$ increases to the next integer power of $(1+\eps)$. By the
previous discussion on weights, the number of such large updates is at
most $O(\log m/\eps^2)$ per edge.  This amortized efficiency comes at
the expense of approximating the (true) weight of each edge to a
$(1+O(\eps))$-multiplicative factor. Happily, an approximate minimum
cut on approximate edge weights is still an approximate minimum cut to
the true weights, so these approximations are tolerable.  In
\citep{cq-17} we isolated a specific lazy-weight scheme from
\citet{y-14} into a data structure with a clean interface that we
build upon here.

\paragraph{Obtaining a primal solution.} Our MWU algorithm computes a
$(1-\eps)$-approximate solution to \ecssd. Recall that the algorithm
maintains a weight $w(e)$ for each edge $e$. Standard arguments show
that one can recover from the evolving weights a
$(1+\eps)$-approximate solution to the LP \ecss (which is the dual of
2ECCSD). A sketch is provided in the appendix.

\section{Tree packings and epochs}
\labelsection{tree-packings}

\citet{k-00} gave a randomized algorithm that finds the global minimum
cut in a weighted and undirected graph with high probability in
$\apxO{m}$ time. Treating Karger's algorithm as a black box in every
iteration leads to a quadratic running time of $\apxO{m^2 / \eps^2}$,
so we open it up and adjust the techniques to our setting.

\paragraph{Tree packings.} \citet{k-00} departs from previous
algorithms for minimum cut with an approach based on packing spanning
trees. Let $\defgraph$ be an undirected graph with positive edge
weights $\weight: \edges \to \preals$, and let
$\trees \subseteq \subsetsof{\edges}$ denote the family of spanning
trees in $\graph$. A \emph{tree packing} is a nonnegatively weighted
collection of spanning trees, $\packing: \trees \to \nnreals$, such
that for any edge $e$, the total weight of trees containing $e$ is at
most $\capacity{e}$ (i.e.,
$\sum_{T \ni e} \packing(T) \leq \capacity{e}$).  Classical work of
\citet{t-61} and \citet{n-61} gives an exact characterization for the
value of a maximum tree packing in a graph, which as an easy corollary
implies it is at least half of the value of a mincut.  If
$C \in \cuts$ is a minimum cut and $\packing$ is a maximum packing,
then a tree $T \in \trees$ selected randomly in proportion to its
weight in the packing will share $\leq 2$ edges with $C$ in
expectation. By Markov's inequality, $T$ has strictly less than $3$
edges in $C$ with constant probability.

For a fixed spanning tree $T$, a \emph{one-cut} in $G$ induced by an
edge $e \in T$ is the cut $\cut{X}$ where $X$ is the vertex set of one
of the components of $T - e$. A \emph{two-cut} induced by two edges
$e_1,e_2 \in T$ is the following.  Let $X,Y,Z$ be the vertex sets of
the three components of $T-\{e_1,e_2\}$ where $X$ is only incident to
$e_1$ and $Z$ is only incident to $e_2$ and $Y$ is incident to both
$e_1,e_2$. Then the two-cut induced by $e_1,e_2$ is
$\cut{Y} = \cut{X \cup Z}$.  Thus, if $T$ is a spanning tree sampled
from a maximum tree packing, and $C$ is a minimum cut, then $C$ is
either a one-cut or a two-cut with constant probability.

The probabilistic argument extends immediately to approximations.  Let
$\zeta \in (0,1)$ be sufficiently small. If $C$ is an approximately
minimum cut with $\sumweight{C} \leq (1+\zeta) \kappa$, and $\packing$
is a tree packing of total weight $\geq (1-\zeta) \kappa/2$, then a
random tree $T$ sampled from $\packing$ has $2 + O(\zeta)$ edges from
$C$ in expectation, and strictly less than three edges in $T$ with
constant probability. Sampling $O(\log (1/\delta))$ trees from a tree
packing amplifies the probability of a tree containing $\leq 2$ edges
of $C$ from a constant to $\geq 1 - \delta$.  For constant
$\zeta > 0$, a $(1-\zeta)$-approximate tree-packing can be computed in
$\apxO{m}$ time, either by applying $\apxO{\kappa m}$-time tree
packing algorithms \citep{g-95,pst-95} to a randomly sparsified graph
\citep{k-98}, or directly and deterministically in $\apxO{m}$ time by
a recent algorithm of \citet{cq-17}.

Another consequence of \citet{k-00} is that for $\zeta < 1/2$, the
number of $(1+\zeta)$-approximate minimum cuts is at most $n^2$. By
the union bound, if we select $O(\log n)$ trees at random from an
approximately maximum tree packing, then with high probability
\emph{every} $(1+\zeta)$-approximate minimum cut is induced by one or
two edges in one of the selected trees.  In summary, we have the
following.
\begin{theorem}[{\citealp{k-00}}]
  \labeltheorem{karger-tree-packing} Let $\defgraph$ be an undirected
  graph with edge capacities $\capacity: \edges \to \preals$, let
  $\delta \in (0,1)$, and let $\eps \in (0,1/2)$. One can generate, in
  $\bigO{m + n \log^3 n}$ time, $h = O(\logof{n/\delta})$ spanning
  trees $T_1,T_2,\dots,T_h$ such that with probability
  $\geq 1 - \delta$, \emph{every} $(1 + \eps)$-approximate minimum cut
  $C$ has $\sizeof{C \cap T_i} \leq 2$ for some tree $T_i$.
\end{theorem}

\citet{k-00} finds the minimum cut by checking, for each tree $T_i$,
the minimum cut in $G$ obtained by removing one or two edges from
$T_i$ in $\apxO{m}$ time. (The details of this subroutine are reviewed
in the following section.) Since there are $O(\log n)$ trees, this
amounts to a near-linear-time algorithm. Whereas Karger's algorithm
finds one minimum cut, we need to find many minimum cuts. Moreover,
each time we find one minimum cut, the edge weights on that cut
increase and so the underlying graph changes. We are faced with the
challenge of outputting $\apxO{m/\eps^2}$ cuts from a dynamically
changing graph that in particular adapts to each cut output, all in
roughly the same amount of time as a single execution of Karger's
algorithm.

\paragraph{Epochs.} We divide the problem into $O(\log m/\eps^2)$
epochs. Recall that at the start of an epoch we have a target value
$\lambda > 0$ and are guaranteed that there are no cuts with value
strictly less than $\lambda$. Our goal is to repeatedly output cuts
with value $\leq (1 + O(\eps))\lambda$ or certify that there are no
cuts of value strictly less than $(1+\eps) \lambda$, in $\apxO{m}$
time. (The reason we output a cut with value $\leq (1+O(\eps))\lambda$
is because the edge weights are maintained only approximately.)  Each
epoch is (conceptually) structured as follows.

\begin{itemize}
\item At the start of the epoch we invoke
  \reftheorem{karger-tree-packing} to find $h = O(\logof{n/\eps})$
  trees $T_1,\ldots, T_h$ such that with probability
  $1 - \poly{\eps/n}$, every $(1+\eps)$-approximate mincut of $G$ with
  respect to the weights at the start of the epoch is a one-cut or
  two-cut of some tree $T_i$.
\item Let $C_1,C_2,\ldots,C_r$ be an {\em arbitrary} enumeration of
  one and two-cuts induced by the trees.
\item For each such cut $C_j$ in the order, if the {\em current
    weight} of $C_j$ is $\le (1+O(\eps)) \lambda$ output it to the MWU
  algorithm and update weights per the MWU framework (weights only
  increase).  Reuse $C_j$ as long as its current weight is
  $\le (1+O(\eps)) \lambda$.
\end{itemize}

Since weights of cuts only increase, and we consider every one and
two-cut of the trees we obtain the following lemma.
\begin{lemma}
  \labellemma{epoch-weight-increase} Let $w: \edges \to \preals$ be
  the weights at the start of the epoch and suppose mincut of $G$ with
  respect to $w$ is $\ge \lambda$; and suppose the trees
  $T_1,\ldots,T_h$ have the property that every $(1+\eps)$-approximate
  mincut is induced by a one or two-edge cut of one of the trees. Let
  $w': \edges \to \preals$ be the weights at the end of an epoch.
  Then, the mincut with respect to weights $w'$ is
  $\ge (1+\eps)\lambda$.
\end{lemma}

\begin{figure}[t]
  \input{algos/held-karp-trees}
  \caption{A third implementation of the MWU framework applied to
    packing cuts that samples a tree packing once at the beginning of
    each epoch and searches for approximately minimum cuts induced by
    one or two edges in the sampled
    trees. \labelfigure{held-karp-trees}}

  \hrulefill
\end{figure}

Within an epoch, we must consider all cuts induced by 1 or 2 edges in
$\apxO{1}$ spanning trees. We have complete flexibility in the order
we consider them. For an efficient implementation we process the trees
one at a time. For each tree $T$, we need to repeatedly output cuts
with value $\leq (1 + O(\eps))\lambda$ or certify that there are no
cuts of value strictly less than $(1+\eps) \lambda$ induced by one or
two edges of $T$. Although the underlying graph changes, we do not
have to reconsider the same tree again because the cut values are
non-decreasing. If each tree can be processed in $\apxO{m}$ time, then
an entire epoch can be processed in $\apxO{m}$ time and the entire
algorithm will run in $\apxO{m/\eps^2}$ time. In
\reffigure{held-karp-trees}, we revise the algorithm to invoke
\reftheorem{karger-tree-packing} once per epoch and use the trees to
look for approximately minimum cuts, while abstracting out the search
for approximately minimum cuts in each tree.

\section{Cuts induced by a tree}
\labelsection{tree-cuts}

Let $\tree \in \trees$ be a rooted spanning tree of $\graph$, and let
$\lambda > 0$ be a fixed value such that the minimum cut value w/r/t
$\weight$ is $\geq \lambda$. We want to list cuts $C \in \cuts$
induced by 1 or 2 edges in $\tree$ with weight
$\sumweight{C} \leq (1+\eps)\lambda$, in any order, but with a twist:
each time we select one cut $C$, we increment the weight of every edge
in $C$, and these increments must be reflected in all subsequent
cuts. We are allowed to output cuts of value
$\leq (1 + O(\eps)) \lambda$. When we finish processing $T$, we must
be confident that there are no 1-cuts or 2-cuts $C \in \cuts_T$
induced by $\tree$ with weight $\sumweight{C} \leq (1+\eps)\lambda$.

The algorithmic challenge here is twofold. First, we need to find a
good 1-cut or 2-cut in $\tree$ quickly. Second, once a good cut is
found, we have to increment the weights of all edges in the
cut. Either operation, if implemented in time proportional to the
number of edges in the cut, can take $\bigO{m}$ time in the worst
case. To achieve a nearly linear running time, both operations must be
implemented in polylogarithmic amortized time.

\begin{figure}[t]
  \begin{ALGO}
    Lazy Incremental Cuts: Interface
  \end{ALGO}

  \begin{apidesc}
    \OPERATIONSDESC %
    \refroutine{lazy-inc-cuts} %
    & %
    Given an undirected graph $\defgraph$ with capacities
    $\capacity: \edges \to \preals$ and edge weights
    $\weight: \edges \to \preals$, and a rooted spanning tree $\tree$,
    initializes the \refroutine{lazy-inc-cuts} data structure.
    \\
    \refroutine{inc-cut} %
    & %
    Given a 1-cut or 2-cut $C \in \cuts_T$ induced by $T$, simulates a
    weight increment along the cut per the MWU framework. Returns a
    list of tuples
    $\Delta = \setof{(e_1,\delta_1),(e_2,\delta_2),\dots}$, where each
    $e_i \in \edges$ is an edge and each $\delta_i > 0$ is a positive
    real value that signals that the weight of the edge $e_i$ has
    increased by an additive factor of $\delta_i$. Each increment
    $(e_i,\delta_i)$ satisfies
    $\delta_i \geq \bigOmega{\eps \weight{e} / \log^2 n}$.
    \\
    \refroutine{flush}{} & %
    Returns a list of tuples
    $\Delta = \setof{(e_1,\delta_1),\dots,(e_k,\delta_k)}$ of all
    residual weight increments not accounted for in previous calls to
    \refroutine{inc-cut} and \refroutine{flush}.
  \end{apidesc}
  \caption{The interface for the data structure
    \refroutine{lazy-inc-cuts}, which efficiently simulates weight
    increments along cuts induced by 1 or 2 edges in a fixed and
    rooted spanning tree. Implementation details are provided later in
    \refsection{lazy-cut-weights}. \labelfigure{lazy-inc-cuts-api}}

  \hrulefill
\end{figure}

In this section we focus on finding the cuts, and assume that we can
update the edge weights along any 1 or 2-cut efficiently via the
\refroutine{lazy-inc-cuts} data structure. Since the weights are
dynamically changing, we need to describe how the algorithm finding
small 1-cuts and 2-cut in $T$ interacts with the
\refroutine{lazy-inc-cuts} data structure.

\paragraph{Interacting with the lazy weight update data structure:}
The interface for \refroutine{lazy-inc-cuts} is given in
\reffigure{lazy-inc-cuts-api}. The primary function of the
\refroutine{lazy-inc-cuts} data structure is \refroutine{inc-cut},
which takes as input one or two edges in $\tree$, and simulates a
weight update along the corresponding 1-cut or
2-cut. \refroutine{inc-cut} returns a list of weight increments
$\parof{e_i,\delta_i}$, where each $e_i$ is an edge and $\delta_i > 0$
is a positive increment that we actually apply to the underlying
graph. The cumulative edge weight increments returned by the
subroutine will always underestimate the true increase, but by at most
a $(1 + O(\eps))$-multiplicative factor for every edge. That is, if we
let $\weight{e}$ denote the ``true weight'' of an edge $e$ implied by
a sequence of weight increments along cuts, and $\apxweight{e}$ the
sum of $\delta$ increments for $e$ returned by \refroutine{inc-cut},
then the \refroutine{lazy-inc-cuts} data structure maintains the
invariant that
\begin{math}
  \apxweight{e} \leq \weight{e} \leq (1 + \rho \eps) \apxweight{e},
\end{math}
where $\rho$ is a fixed constant that we can adjust.  In particular, a
cut of weight $\leq (1 + \eps) \lambda$ w/r/t the approximated edge
weights retains weight $\leq (1+O(\eps)) \lambda$ in the true
underlying graph.  Each increment $(e,\delta)$ returned by
\refroutine{inc-cut} reflects a
$(1 + \Omega(\eps / \log^2 n))$-multiplicative increase to
$\weight{e}$.  As the weight of an edge is bounded above by
$\apxO{m^{1/\eps}}$, the subroutine returns an increment for a fixed
edge $e$ at most $\bigO{\log^3 n/\eps^2}$ times total over the entire
algorithm.

A second routine, \refroutine{flush}{}, returns a list of increments
that captures all residual weight increments not accounted for in
previous calls to \refroutine{inc-cut} and \refroutine{flush}. The
increments returned by \refroutine{flush}{} may not be very large, but
restores $\apxw{e} = \weight{e}$ for all $e$. Formally, we will refer
to the following.

\begin{lemma}
  \labellemma{lazy-inc-cuts}

  Let $\defgraph$ be an undirected graph with positive edge capacities
  $\capacity: \edges \to \preals$ and positive edge weights
  $\weight: \edges \to \preals$, and $T$ a rooted spanning tree in
  $\graph$. Consider an instance of
  \refroutine{lazy-inc-cuts}{\graph}{\capacity}{\weight}{T} over a
  sequence of calls to \refroutine{inc-cut} and
  \refroutine{flush}. For each edge $e$, let $\weight{e}$ be the
  ``true'' weight of edge $e$ if the MWU framework were followed
  exactly, and let $\apxweight{e}$ be the approximate weight implied
  by the weight increments returned by calls to \refroutine{inc-cut}
  and \refroutine{flush}.
  \begin{mathresults}
  \item After each call to \refroutine{inc-cut}, we have
    $\apxw{e} \leq \weight{e} \leq (1 + O(\eps)) \apxw{e}$ for all
    edges $e \in \edges$.
  \item After each call to \refroutine{flush}, we have
    $\apxw{e} = \weight{e}$.
  \item Each call to \refroutine{inc-cut} takes $O(\log^2 n + I)$,
    where $I$ is the total number of weight increments returned by
    \refroutine{inc-cut}.
  \item Each call to \refroutine{flush} takes $O(m \log^2 n)$ time and
    returns at most $m$ weight increments.
  \item \labelitem{lazy-inc-cuts-min-increment} Each increment
    $(e,\delta)$ returned by \refroutine{inc-cut} satisfies
    $\delta \geq \bigOmega{ (\eps / \log^2 n) \weight{e} %
    }$.
  \end{mathresults}
\end{lemma}

\begin{figure}[t!]
  \input{algos/held-karp-lazy-cuts} %
  \caption{A fourth implementation of the MWU framework applied to the
    Held-Karp relaxation, that integrates the
    \refroutine{lazy-inc-cuts} data structure into the subroutine that
    processes each tree. \labelfigure{held-karp-lazy-inc-cuts}}

  \hrulefill
\end{figure}

The implementation details and analysis of \refroutine{lazy-inc-cuts}
are given afterwards in \refsection{lazy-cut-weights}.  An enhanced
sketch of the algorithm so far, integrating the
\refroutine{lazy-inc-cuts} data structure but abstracting out the
search for small 1- and 2-cuts, is given in
\reffigure{held-karp-lazy-inc-cuts}.  In this section, we assume
\reflemma{lazy-inc-cuts} and prove the following.
\begin{lemma}
  \labellemma{cut-search} %
  Let $\tree$ be a rooted spanning tree and $\lambda > 0$ a target cut
  value. One can repeatedly output 1-cuts and 2-cuts $C \in \cuts_T$
  induced by $T$ of weight $\sumweight{C} \leq (1 + O(\eps)) \lambda$
  and increment the corresponding edge weights (per the MWU framework)
  until certifying that there are no 1-cuts or 2-cuts $C \in \cuts_T$
  of value $\sumweight{C} \leq (1 + \eps) \lambda$ in
  $\bigO{m \log^2 n + K \log^2 n + I \log n}$ time, where $K$ is the
  number of 1-cuts and 2-cuts of value $\leq (1 + O(\eps)) \lambda$
  output and $I$ is the total number of weight increments returned by
  the \refroutine{lazy-inc-cuts} data structure.
\end{lemma}

The rest of the section is devoted to the proof of the preceding
lemma.  We first set up some notation.  Fixing a rooted spanning tree
$\tree$ induces a partial order on the vertices $\vertices$. For two
vertices $u$ and $v$, we write $u \leq v$ if $u$ is a descendant of
$v$.  Two vertices $u$ and $v$ are \emph{incomparable}, written
$u \incomp v$, if neither descends from the other.  For each vertex
$v \in \vertices$, let $\subtree{v}$ be the subtree of $\tree$ rooted
at $v$. Let
\begin{math} D(v) \defeq \vertices(\subtree{v}) = \setof{u: u \leq v}
\end{math} be the \emph{down set} of all descendants of $v$ in $\tree$
(including $v$). We work on a graph with edge weights given by a
vector $\weight: \edges \to \preals$.  For a  set of vertices
$S \subset \vertices$, we let
\begin{math}
  \cut{S} \defeq \setof{e \in \edges \suchthat \sizeof{e \cap S} = 1}
\end{math} be the set of edges cut by $S$, and let
\begin{math}
  \cutw{S}\defeq \sumweight{\cut{S}} = \sum_{e \in \cut{S}} \weight{e}
\end{math} denote the weight of the cut induced by $S$. For two
disjoint sets of vertices $S,T \subseteq \vertices$, we let
\begin{math}
  \cut{S,T} = \setof{e = (a,b) \in \edges \suchthat a \in S, b \in T}
\end{math} be the edges crossing from $S$ to $T$, and let
$\cutw{S,T} = \sumweight{\cut{S,T}}$ be their sum weight.

Per \citet{k-00}, we divide the 1-cuts and 2-cuts into three distinct
types, drawn in \reffigure{tree-cuts}. First there are the 1-cuts of
$\tree$ of the form $\cut{D(s)}$. Then we have 2-cuts
$\cut{D(s) \cup D(t)}$ where $s$ and $t$ are incomparable (i.e.,
$s \incomp t$). We call these \emph{incomparable 2-cuts}. Lastly, we
have 2-cuts
\begin{math} \cut{D(t) \setminus D(s)}
\end{math} where $s$ is a descendant of $t$ (i.e., $s < t$). We call
these \emph{nested 2-cuts}. For a fixed tree, we process all 1-cuts,
incomparable 2-cuts, and then nested 2-cuts, in this order. For each
of these three cases we prove a lemma similar to \reflemma{cut-search}
but restricted to the particular type of cut. Throwing in $\apxO{m}$
time to initialize the \refroutine{lazy-inc-cuts} data structure at
the beginning and $\apxO{m}$ to call \refroutine{flush} at the end
then gives \reflemma{cut-search}.

\begin{figure}
  \hfill%
  \begin{minipage}{.25\textwidth}
    \centering%
    \includegraphics[width=\textwidth]{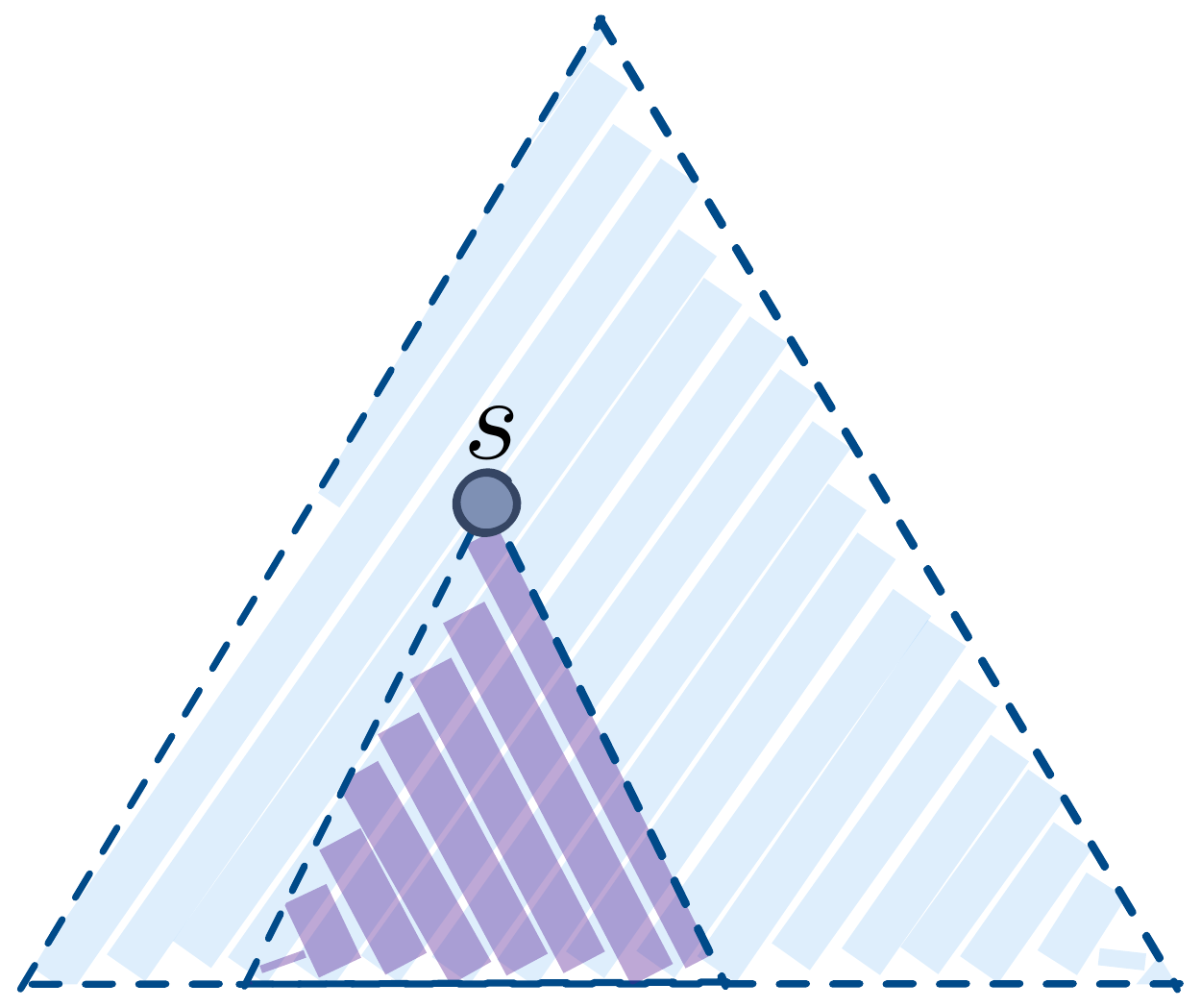}

    \smallskip %
    \emph{1-cut}
  \end{minipage}%
  \hfill%
  \begin{minipage}{.25\textwidth}
    \centering%
    \includegraphics[width=\textwidth]{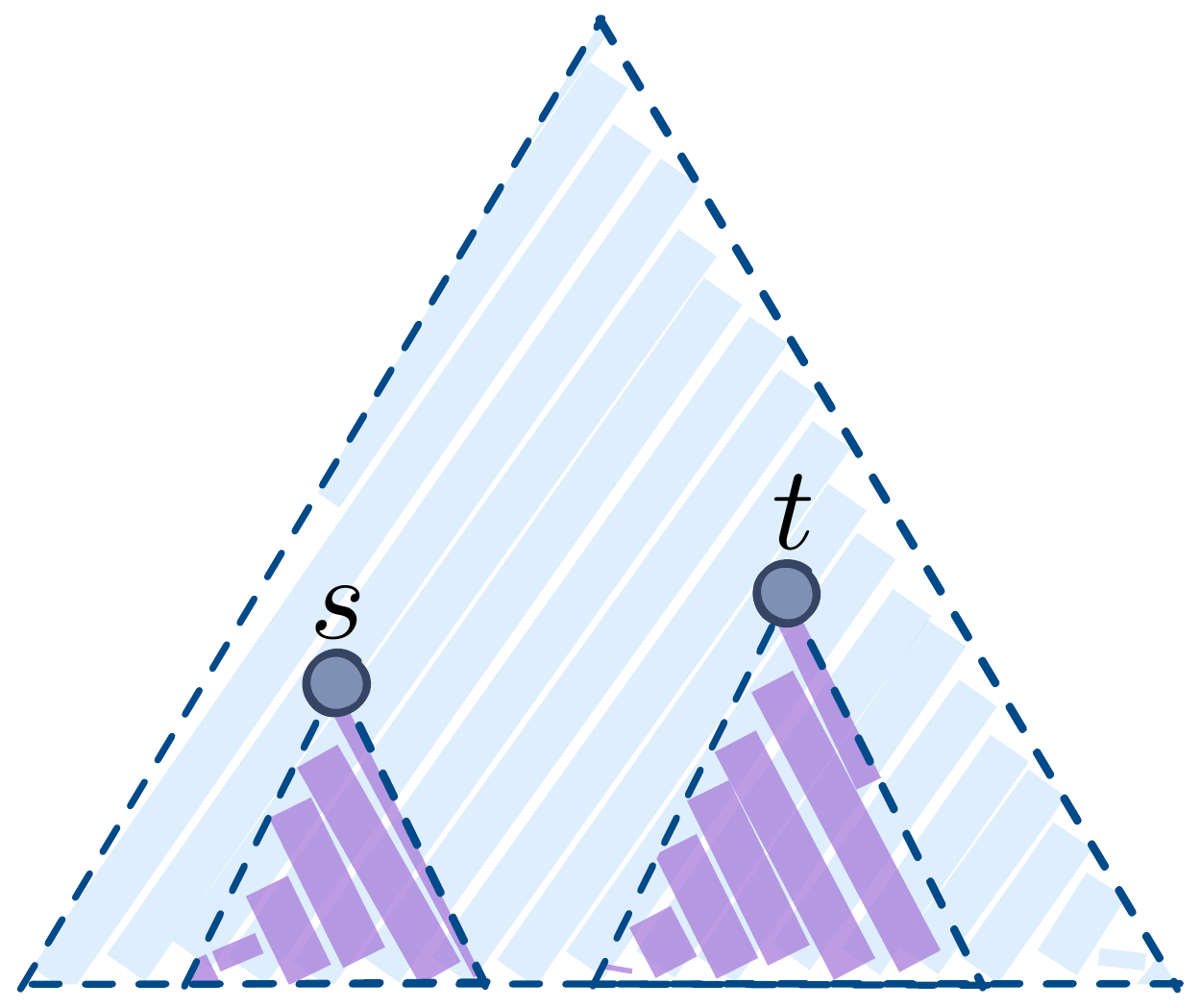} %

    \smallskip \emph{Incomparable 2-cut}
  \end{minipage}%
  \hfill%
  \begin{minipage}{.25\textwidth}
    \centering %
    \includegraphics[width=\textwidth]{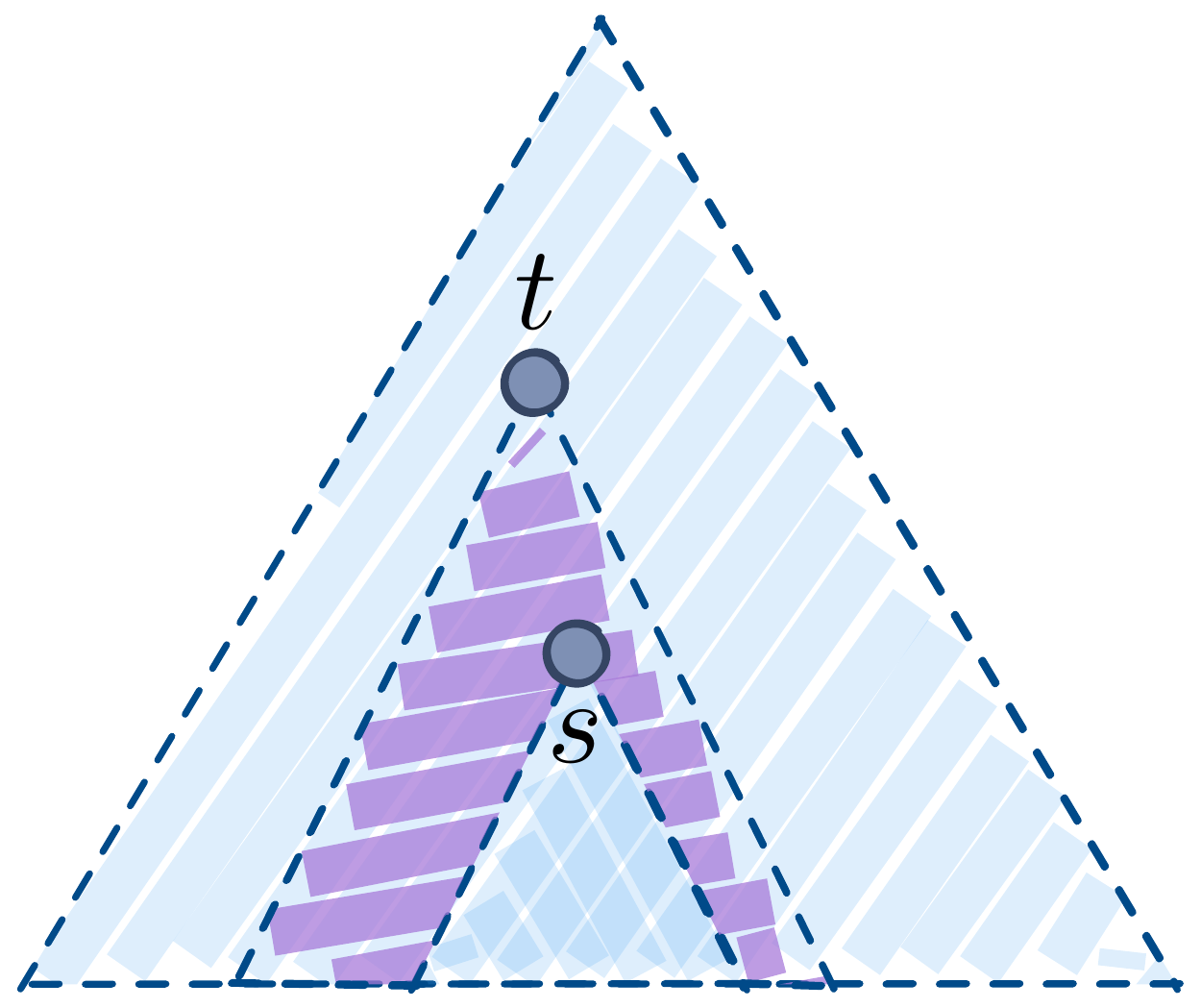} %

    \smallskip \emph{Nested 2-cut} %
  \end{minipage}%
  \hfill\hfill%

  \caption{3 types of tree-cuts \labelfigure{tree-cuts}}
\end{figure}

\paragraph{Standard data structures.} Following \citet{k-00}, we rely
on the dynamic tree data structure, \emph{link-cut trees}, of
\citet{st-83}. Link-cut trees store real values at the nodes of a
rooted tree $T$ and support aggregate operations along node-to-root
paths in logarithmic time. We employ two operations in
particular. First, given a node $v$ and a value $\alpha \in \reals$,
we can add $\alpha$ to the value of every node on the $v$-to-root path
(i.e., every node $u$ such that $u \geq v$) in $\bigO{\log n}$
amortized time. Second, given a node $v$, we can find the minimum
value of any node on the $v$-to-root path in $\bigO{\log n}$ amortized
time.

We also need to compute least common ancestors ($\lca$) in $T$ for
pairs of nodes. \citet{ht-84} showed how to preprocess $T$ in linear
time and answer $\lca$ queries in constant time.

\subsection{1-cuts}

\labelsection{1-cuts} %

Let $e = (t,s) \in \tree$ be a tree edge with $t$ a parent of
$s$. Then $\tree - e$ consists of two components: the subtree
$\subtree{s}$ rooted at $s$, and the remaining tree
$T \setminus \subtree{s}$. The cut $\cut{D(s)}$ in the underlying
graph $\graph$ induced by the subtree $\subtree{s}$ is a ``1-cut'' of
$\tree$. In this section, we want to output a sequence of 1-cuts
$C \in \cuts_T$ in $\tree$ with
$\sumweight{C} \leq (1+O(\eps))\lambda$, repeating if necessary, while
incrementing weights of each listed cut on the fly, until we've
determined that there are no 1-cuts $C$ with weight
$\sumweight{C} \leq (1+\eps) \lambda$.

\paragraph{Computing and maintaining the 1-cut values.}
To find the minimum 1-cut induced by a down set $D(v)$, Karger
observes that the weight $\cutw{\downset{v}}$ can be rewritten as
\begin{align*}
  \cutw{\downset{v}}
  =                             %
  \sum_{x \in D(v)} \cutw{x} - 2 \sum_{e \in \edges[D(v)]} \weight{e},
\end{align*}
where $\cutw{x}$ is the weighted degree of the vertex $x$, and
$\edges[D(v)]$ is the set of edges with both endpoints in $D(v)$.

The quantities
\begin{math}
  \sum_{x \in D(v)} \cutw{x}
\end{math}
over $v \in \vertices$ are just tree sums over the weighted degrees,
and \citet{k-00} computes them once for all $v$ in $O(m)$ total time
by a depth-first traversal. In the dynamic setting, we can compute and
maintain the tree sums in $\bigO{\log n}$ time per edge update in a
link-cut tree. Each time we increment the value of an edge
$e = (u,v) \in \edges$ by some $\delta > 0$, we add $\delta$ to the
value of every node on the paths to the root from $u$ and $v$.

The sum
\begin{math}
  \sumweight{\edges[D(v)]} \defeq \sum_{e \in \edges[D(v)]} \weight{e}
\end{math}
over edges with both endpoints in $D(v)$ is just the sum of weights of
all edges $e = (x,y)$ for which the least common ancestor $\lca{x,y}$
of $x$ and $y$ is in $D(v)$; i.e.,
\begin{math}
  \sumweight{\edges[D(v)]} %
  = %
  \sum_{e \suchthat \lca{e} \in D(v)} \weight{e}
\end{math}
By \citet{ht-84}, we can compute $\lca{e}$ for every edge
$e \in \edges$ in linear time. We then compute for each vertex $v$ the
sum $\sum_{\lca{e} = v} \weight{e}$ of all weights of edges whose
least common ancestor is $v$. The quantities
$\sumweight{\edges[(D(v))]}$ can then be computed as tree-sums by
depth-first search, and updated dynamically with link-cut trees,
similar to before.

The preceding discussion shows that the sums $\cutw{\downset{v}}$ for
each $v$ can be maintained dynamically in logarithmic time using
link-cut trees. This is summarized in the lemma below.

\begin{lemma}
  \labellemma{update-1-cut} Given edge-weighted graph $\defgraph$ and
  a rooted spanning tree $T$, there is a data structure to maintain
  $\cutw{D(v)}$ for all $v \in \vertices$ as edge-weights are changed.
  The initialization cost is $O(m)$, updates and queries take
  $\bigO{\log n}$ amortized time.
\end{lemma}

\paragraph{Processing all 1-cuts.} The weight of all 1-cuts can be
computed in $\bigO{m}$ time statically and $\bigO{\log n}$-time per
edge update. To process all 1-cuts of the tree, we consider all
downsets $D(v)$ in any order. For each vertex $v$, we check if
$\cutw{D(v)} \leq (1+\eps)\lambda$ in $\bigO{\log n}$ amortized
time. If not, then we move on to the next vertex. Since the edge
weights are monotonically increasing, $\cutw{D(v)}$ will never be
$\leq (1+\eps)\lambda$ again, so we do not revisit $v$.  Otherwise, if
$\cut{D(v)}$ is good, we take the cut in our solution and pass $v$ to
\refroutine{inc-cut} to signal a weight increment along
$\cut{D(v)}$. In turn, \refroutine{inc-cut} returns a list of
``official'' edge increments $\setof{(e_i,\delta_i)}$.  Each update
can be incorporated into the link-cut tree in logarithmic time by
\reflemma{update-1-cut}. After processing the edge increments returned
by \refroutine{inc-cut}, we continue to process $D(v)$ until
$\cut{D(v)}$ is no longer a good cut. The total running time to
process all 1-cuts is $\bigO{m}$ plus $\bigO{\log^2 n}$ per cut output
and $\bigO{\log n}$ per edge weight increment returned by
\refroutine{inc-cut}.

\begin{lemma}
  Let $T$ be a fixed and rooted spanning tree and $\lambda > 0$ a
  fixed target cut value. Employing the \refroutine{inc-cut} routine
  of the \refroutine{lazy-inc-cuts} data structure to approximately
  increment edge weights, one can repeatedly find 1-cuts $C$ induced
  by $T$ of value $\cutweight{C} \leq (1 + O(\eps)) \lambda$ and
  increment the corresponding edge weights (per the MWU framework)
  until certifying that there are no 1-cuts $C$ of value
  $\cutweight{C} \leq (1 + \eps) \lambda$ in
  $\bigO{m + K \log^2 n + I \log n}$ total running time, where $K$ is
  the number of 1-cuts of value $\leq (1 + O(\eps))$ output and $I$ is
  the total number of weight increments returned by
  \refroutine{inc-cut}.
\end{lemma}

\subsection{2-cuts on incomparable vertices}

In this section, we consider incomparable 2-cuts of the form
$\cut{D(s) \cup D(t)}$, where $s \incomp t$.  We first consider the
case where $s$ is a fixed leaf, and $t$ ranges over all vertices
incomparable to $s$. We then extend the leaf case to consider the case
where $s$ ranges over a path in the tree down to a leaf, and $t$
ranges over all incomparable vertices to the path. We then apply an
induction step that reduces processing the whole tree to a logarithmic
number of rounds where in each round we process all the paths to
leaves. See \reffigure{incomparable-2-cut-cases}.

\begin{figure}
  \hfill%
  \includegraphics[width=.2\textwidth,valign=t]{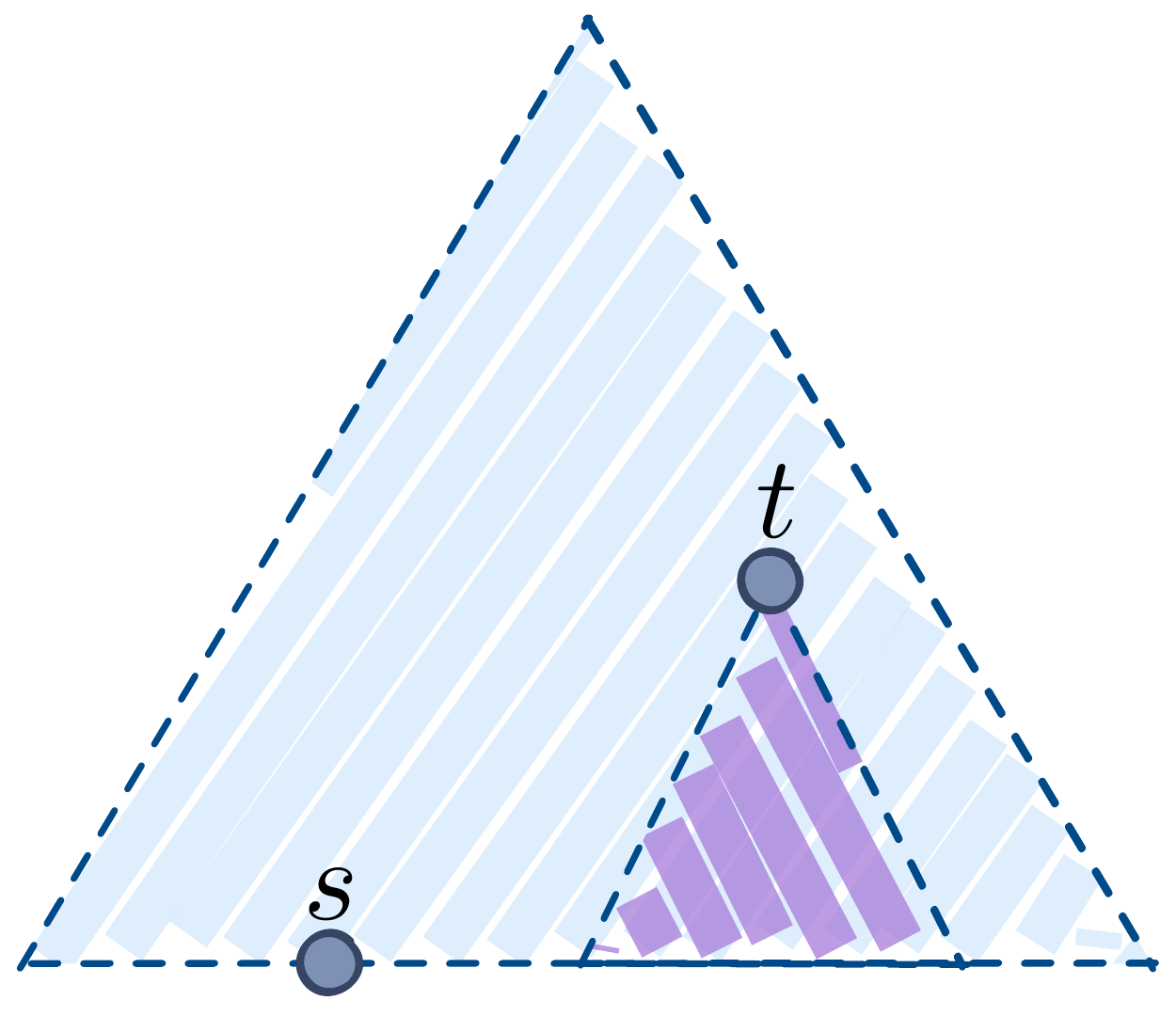}
  \hfill%
  \includegraphics[width=.2\textwidth,valign=t]{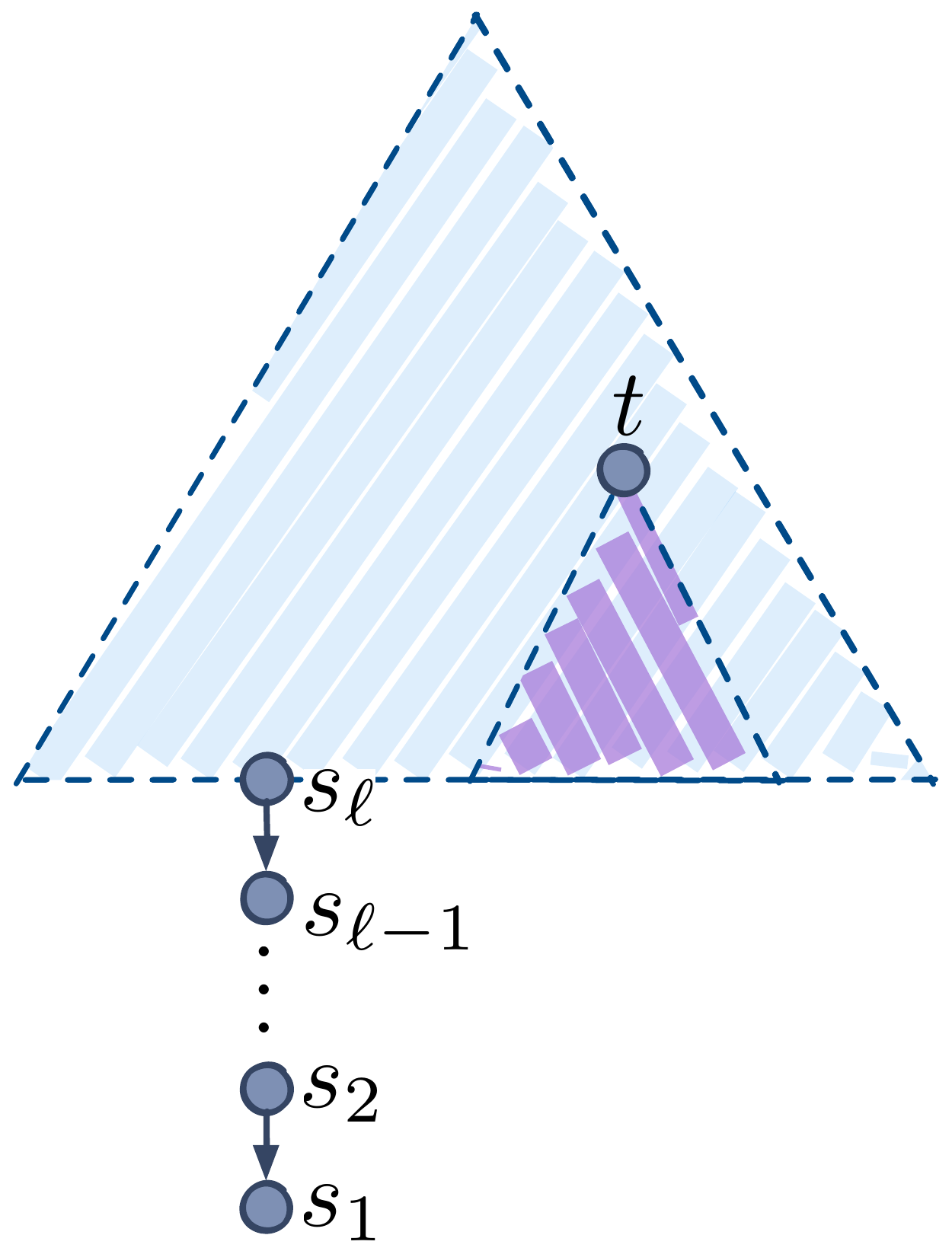} %
  \hfill%
  \includegraphics[width=.2\textwidth,valign=t]{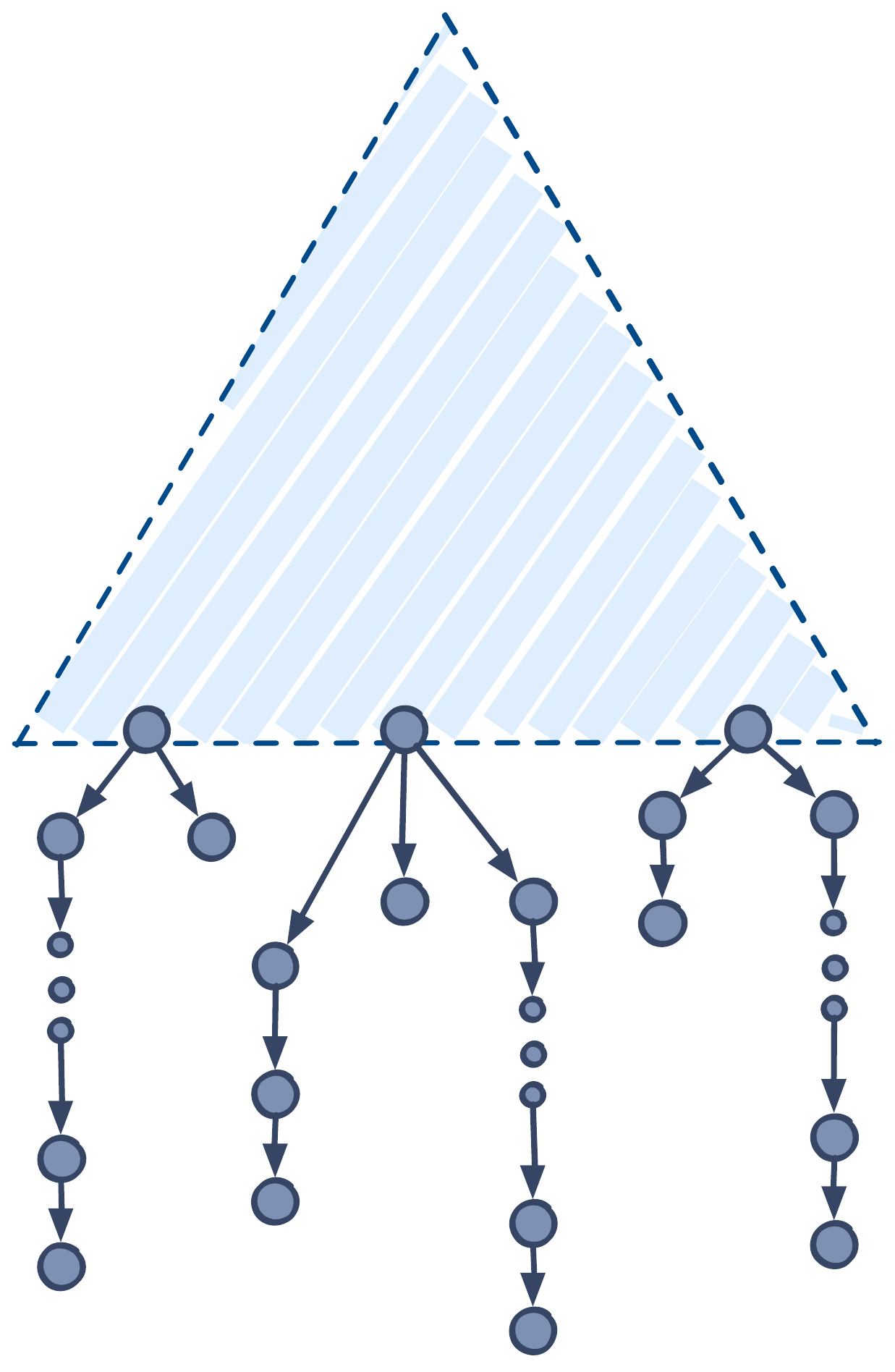} %
  \hfill\hfill%

  \hfill%
  \begin{minipage}[t]{.2\textwidth}
    \centering%
    \emph{Fixed leaf}
  \end{minipage}%
  \hfill%
  \begin{minipage}[t]{.2\textwidth}
    \centering%
    \emph{Fixed path to a leaf}
  \end{minipage}%
  \hfill%
  \begin{minipage}[t]{.2\textwidth}
    \centering %
    \emph{Induction step} %
  \end{minipage}%
  \hfill\hfill%

  \caption{3 cases for incomparable 2-cuts
    \labelfigure{incomparable-2-cut-cases}}

\end{figure}

\paragraph{Incomparable 2-cuts with a leaf.} We first fix $s$ to be a
leaf in $T$, and consider all incomparable 2-cuts with one side fixed
to be $\setof{s}$.  Karger observes that the weight of the cut induced
by $D(t) + s$, for $t \incomp s$, can be written as
\begin{align*}
  \cutweight{D(t) + s}
  =                             %
  \cutweight{s} + \cutweight{D(t)} - 2\cutweight{D(t),s}
\end{align*}
where $\cutweight(D(t),s)$ is the weight of edges with one endpoint
$s$ and the other in $D(t)$.  We only need to consider down sets
$D(t)$ for which $s$ and $D(t)$ are connected by some edge, since
otherwise either $\setof{s}$ or $D(t)$ induce a smaller cut, and after
processing all 1-cuts, neither $\setof{s}$ or $D(t)$ induce small
enough cuts.

For fixed $s$ and ranging $t$ over all $t \incomp s$, Karger finds the
weight of the best incomparable $D(t) + s$ cut statically as follows.
As $\cutweight{s}$ is fixed, the $t$'s are differentiated only by the
values
\begin{math}
  \cutweight{D(t)} - 2 \cutweight{D(t),s}.
\end{math}
Karger's algorithm creates a dynamic tree over $\tree$ with each
vertex $t$ initialized with the value $\cutweight{D(t)}$. (This
initial tree only needs to be constructed once, and immutably reused
for other $s$'s). Karger's algorithm immediately adds $+\infty$ to the
value of every vertex on the $s$-to-root path to eliminate all
comparable vertices from consideration. For each edge $e = (s,v)$ with
endpoint $s$, Karger's algorithm subtracts $2 \weight{e}$ from every
vertex on the $v$-to-root path, to incorporate the second term
$- 2 \cutweight{D(t),s}$. Then, for each edge $e = (s,v)$ adjacent to
$s$, Karger's algorithm finds the minimum value vertex on the $v$ to
root path. The minimum value vertex $z$ over all such queries gives
the best $s + D(z)$ cut subject to $s \incomp z$, as desired. The
running time to process the leaf $s$, excluding the construction of
the initial tree, is bounded by the time spent subtracting edge
weights and taking minimums over the $v$-to-root path for each edge
$(s,v)$ incident to $s$. With dynamic trees, subtracting and finding
the minimum along a vertex-to-root path each take $\bigO{\log n}$
amortized time. Thus, up to a logarithmic factor, the time spent
processing all incomparable 2-cuts with one side a fixed leaf $s$ is
proportional to the number of edges incident to $s$.

In the dynamic setting, we do the following. We first compute the sum
$\cutweight{D(t)} - 2 \cutweight{D(t),s}$ for all incomparable
$t: t \incomp s$, as above. We then run through the edges $(s,v)$
incident to $s$ in any fixed order, checking for $s + D(t)$ cuts with
$t \geq v$ that have value $\leq (1+\eps)\lambda$ and updating edge
weights along good cuts on the fly, as follows.

Fix an edge $e = (s,v)$ incident to $s$. We process $e$ until we know
that there is no good $(D(t) + s)$-cut for any ancestor $t$ of $v$. We
first find the minimum value vertex on the $v$-to-root path in
$\bigO{\log n}$ amortized time. If the minimum value is
$> (1+\eps) \lambda - \cutweight{D(s)}$, then we move onto the next
edge incident to $s$.  Otherwise, let $t$ be an ancestor of $v$ such
that $s + D(t)$ induces a good cut. We pass the cut $\cut{s + D(t)}$
(in terms of the roots of $s$ and $t$) to \refroutine{inc-cut}, which
simulates a weight update along $\cut{s + D(t)}$ and returns a list of
weight increments $\setof{(e_i,\delta_i)}$. For each such edge
$e = (x,y)$ whose weight is increased by some $\delta > 0$, we add
$\delta$ to every vertex on the $x$-to-root and $y$-to-root paths. We
then subtract $2 \delta$ from the $\lca{x,y}$-to-root path. These
additions and subtraction account for the first term,
$\cutweight{D(t)}$.  If $x = s$, then we also subtract $2\delta$ from
the $y$-to-root path for the sake of the second term,
$-2 \cutweight{D(t),s}$. After all official weight increments returned
by \refroutine{inc-cut} have been incoporated into the dynamic tree,
we continue to process the same edge $e = (s,v)$. Once all edges
$e = (s,v)$ incident to $s$ have been processed, we have certified
that all 2-cuts with one side fixed to the leaf $s$ have weight
$\geq (1 + \eps) \lambda$. The total time is $\bigO{\log n}$ times the
number of edges incident to $s$, plus $\bigO{\log^2 n}$ time for every
cut output and $\bigO{\log n}$ for every increment returned by
\refroutine{inc-cut}.

\paragraph{Leaf-paths.} The approach outlined above for processing
leaves extends to processing paths. Let a \emph{leaf-path} be a
maximal subtree that is a path. Let $p = s_1,s_2,\dots,s_{\ell}$ be a
leaf path, where $s_1$ is a leaf, $s_2$ is its parent, and so
forth. We want to process all 1-cuts of the form $D(s_i) \cup D(t)$,
where $t$ is incomparable to any and all of the $s_i$'s on the path.

For each $i$, we have
\begin{align*}
  \cutw{D(s_i) \cup D(t)}       %
  =                             %
  \cutw{D(s_i)}
  +                             %
  \cutw{D(t)}                   %
  -                             %
  2 \cutw{D(s_i),D(t)},
\end{align*}
where $\cutw{D(s_i),D(t)}$ is the weight of all edges between $D(s_i)$
and $D(t)$.  The values $\cutw{D(t)}$ for
$t \incomp s_1,\dots,s_{\ell}$ and $\cutw{D(s_i)}$ for each $i$ are
easy to compute initially and maintain dynamically, per the earlier
discussion about 1-cuts in \refsection{1-cuts}.  For fixed $i$, the
incomparable $t$ are differentiated by the value
$\cutw{D(t)} - 2\cutw{D(s_i),D(t)}$. The basic idea here is that
$\cutw{D(s_{i+1}),D(t)}$ can be written in terms of the preceding cut
in the form
\begin{math}
  \cutw{D(s_{i+1}),D(t)} %
  = %
  \cutw{D(s_i),D(t)} %
  + %
  \cutw{s_{i+1},D(t)}.
\end{math}
That is, after computing $\cutw{D(s_i),D(t)}$ for all incomparable
$t$, we can keep these values and just incorporate the weights of
edges incident to $s_{i+1}$ to get the values $\cutw{D(s_i),D(t)}$ for
all incomparable $t$. % {\bf Chandra: don't we have to adjust for the
% weight of the edge $s_is_{i+1}$ if there is one?}

We first review the static case, where we want to find the minimum
such cut. Karger's algorithm first processes the leaf $s_1$ as a leaf,
as outlined above, and takes note of the minimum $s_1 + D(t)$ cut. It
then processes $s_2$ in the same fashion, except it continues the
aggregate data built when processing $s_1$ rather than starting
fresh. By subtracting the weights of all edges incident $s_2$ from the
values of the various incomparable downsets, and having already
subtracted the weights of edges incident to $s_1$, we will have
subtracted the weight of all edges incident to
$D(s_2) = \setof{s_1,s_2}$, and so we have computed
$\cutweight{D(t)} - 2 \cutweight{D(s_2),D(t)}$ for all
$t \incomp s_2$.  In this fashion, Karger's algorithm marches up the
path doing essentially the same work for each vertex as for a fixed
leaf, except continuing the same link-cut tree from one $s_i$ to the
next. The total work processing the entire leaf-path is, up to a
logarithmic factor, proportional to the total number of edges incident
to any node $s_i$ on the path.

The analogous adjustments are made in the dynamic setting. That is, we
process $s_1$ as a leaf, passing good cuts to \refroutine{inc-cut} and
incorporating the returned edge increments on the fly. After
processing $s_1$, we keep the accumulated aggregate values, and start
processing $s_2$ likewise.  Marching up the leaf-path one vertex at a
time.  In this manner, we are able to process 2-cuts of the form
$D(s_i) + D(t)$ over each $s_i$ along the leaf-path
$s_1,\dots,s_{\ell}$ (subject to $t \incomp s_1,\dots,s_{\ell}$) and
at the end certify there are no such cuts of weight
$< (1 + \eps) \lambda$. The total work in processing the leaf path is
$\bigO{\log n}$ for each edge incident to $s_1,\dots,s_{\ell}$,
$\bigO{\log^2 n}$ for each good cut output, and $\bigO{\log n}$ for
each weight increment returned from the subroutine
\refroutine{inc-cut}.

\paragraph{All incomparable 2-cuts.}  \citet{k-00} showed that an
efficient subroutine for processing all incomparable 2-cuts on a
single leaf-path leads to an efficient algorithm for processing all
incomparable 2-cuts in the entire tree as follows.  The overall
algorithm processes the 2-cuts of a tree in phases. Each phase
processes all the leaf-paths of the current tree, and then contracts
the leaf-paths into their parents and recurses on the new tree in a
new phase.  Each leaf in the contracted graph had at least two
children in the previous phase, so the number of nodes has
halved. After at most $O(\log n)$ phases, we have processed all the
incomparable 2-cuts in the tree. Excluding the work for selecting a
cut and incrementing edge weights returned by the
\refroutine{lazy-inc-cuts} data structure, processing a leaf-path
takes time proportional to the number of edges incident to the
leaf-path. Each edge is incident to at most 2 leaf-paths. Thus, a
single phase takes $\bigO{m \log n}$ time, in addition to
$\bigO{\log^2 n}$ time for every cut output and $\bigO{\log n}$
amortized time for every edge weight increment returned by the
\refroutine{lazy-inc-cuts} data structure. In conclusion, we have
shown the following.

\begin{lemma}
  Let $T$ be a fixed and rooted spanning tree and $\lambda > 0$ a
  fixed target cut value. Employing the \refroutine{inc-cut} routine
  of the \refroutine{lazy-inc-cuts} data structure to approximately
  increment edge weights, one can repeatedly find incomparable 2-cuts
  $C$ induced by $T$ of value
  $\cutweight{C} \leq (1 + O(\eps))\lambda$ and increment the
  corresponding edge weights (per the MWU framework) until certifying
  that there are no incomparable 2-cuts $C$ of value
  $\cutweight{C} \leq (1 + \eps) \lambda$ in
  $\bigO{m \log^2 n + K \log^2 n + I \log n}$ total amortized time,
  where $K$ is the number of incomparable 2-cuts of value
  $\leq (1 + O(\eps))\lambda$ output and $I$ is the total number of
  weight increments returned by \refroutine{inc-cut}.
\end{lemma}

\subsection{Nested 2-cuts}

The other type of 2-cuts is of the form $D(t) \setminus D(s)$, where
$t$ is an ancestor of $s$. We call these cuts \emph{nested 2-cuts}. As
in the case of incomparable 2-cuts, we first consider the case where
$s$ is fixed to be a leaf, then the case where $s$ is on a leaf-path,
and then finally an induction step that processes all leaf-paths in
each pass.

\paragraph{Nested 2-cuts with a leaf.} We take the same approach as
with incomparable 2-cuts, and start with the case where $s$ is a fixed
leaf. Karger observed that the weight of the cut induced by $D(t) - s$
can be written as
\begin{align*}
  \cutweight{D(t) - s} %
  =                               %
  \cutweight{D(t)} + 2\cutweight{s,D(t)} - \cutweight{s},
  \labelthisequation{nested-2-cut-leaf}
\end{align*}
where $\cutweight{s,D(t)}$ is the sum weight of the edges between $s$
and $D(t)$. We start with a link-cut tree over $\tree$ where each node
$v$ is initialized with $\cutweight{D(v)}$. For each edge $e = (s,u)$
incident to $s$, we find the least common ancestor $t = \lca{s,u}$,
and add $2\weight{e}$ to every node on the $u$-to-root path in
$\bigO{\log n}$ time. We also subtract $\weight{e}$ from the value of
every node on the $s$-to-root path. The first set of updates along the
$t$-to-root path is for the second term of
\refequation{nested-2-cut-leaf}, and the second set along the
$s$-to-root path is for the third term.

Karger's algorithm finds the minimum value on the $s$-to-root path to
get the value of the minimum $(D(t) - s)$-cut over all $t > s$. In the
dynamic setting, we repeatedly find the minimum value on the
$s$-to-root path as long as this value is below the threshold
$(1 + \eps)\lambda$. Each time we find a good cut $\cut{D(t) -s}$, we
send the cut to the routine \refroutine{inc-cut} (compactly
represented by $s$ and $t$). For each edge weight increment
$(e = (x,y),\delta)$ returned by \refroutine{inc-cut} we do the
following. Let $t = \lca{x}{y}$ be the least common ancestor. First,
we add $\delta$ to every the value of every vertex on the $x$-to-root
and $y$-to-root paths and subtract $2 \delta$ from every value of
every vertex on the $t$-to-root-path in $O(\log n)$ time, to account
for the first term $\cutweight{D(t)}$ (see \refsection{1-cuts}). If
$e$ is also incident to $s$, then we add $2 \delta$ to the value of
every vertex on the $t$-to-root path (for the second term of
\refequation{nested-2-cut-leaf}), and subtract $\delta$ from the value
of every vertex on the $s$-to-root path (for the third term). Thus,
incorporating an edge increment $(e,\delta)$ consists of a constant
number of updates along node-to-root paths and takes $\bigO{\log n}$
amortized time.

The total amortized running time for processing all nested 2-cuts of
the form $D(t) - s$ with a fixed leaf $s$ is $\bigO{\log n}$ times the
number of edges incident to $s$, plus $\bigO{\log^2 n}$ for every cut
output and $\bigO{\log n}$ for every edge weight increment returned by
the \refroutine{inc-cut} subroutine.

\paragraph{Nested 2-cuts along a leaf-path.} We extend the leaf case
to a leaf path $s_1,\dots,s_{\ell}$, where $s_1$ is a leaf and each
$s_i$ with $i > 1$ has exactly one child, and consider all cuts of the
form $D(t) \setminus D(s_i)$ over all $s_i$ in the leaf path and all
ancestors $t$ of $s_i$. For each $i$, we have
\begin{align*}
  \cutweight{D(t) \setminus D(s_i)} %
  &=                                 %
    \cutweight{D(t)} + 2\cutweight{D(s_i),D(t)} -
    \cutweight{D(s_i)}. %
    \labelthisequation{nested-cuts-path}
\end{align*}
% \\
% &= %
% \cutweight{D(t)} %
% + %
% 2\sum_{j \leq i} \cutweight{s_j, D(t)} - \cutweight{D(s_i)}.
% \end{align*}
% where $\edges(D(S_i))$ is the set of edges with both endpoints in
% $D(s_i)$.
%
% The first term, $\cutweight{D(t)}$, is just the weight of the 1-cut
% $D(t)$, and easy to compute statically and maintain dynamically per
% the earlier discussion in \refsection{1-cuts}.
When $i = 1$, this is the same as equation
\refequation{nested-2-cut-leaf} obtained for the leaf case.  For
$i = 1,\dots,\ell-1$, the difference in \refequation{nested-cuts-path}
between consecutive vertices $s_i$ and $s_{i+1}$ is
\begin{align*}
  &\cutweight{D(t) \setminus D(s_{i+1})} %
    -                                 %
    \cutweight{D(t) \setminus D(s_{i})} %
  \\
  &=                                    %
    2\bracketsof{                            %
    \cutweight{D(s_{i+1}),D(t)} - \cutweight{D(s_{i}),D(t)} %
    }                                                       %
    -                                                       %
    \bracketsof{\cutweight{D(s_{i+1}) - \cutweight{D(s_{i})}}}
  \\
  &=                            %
    2 \cutweight{s_{i+1}, D(t)}
    +
    \cutweight{s_{i+1}}
    -                           %
    2 \cutweight{s_{i+1},D(s_{i})}.
    \labelthisequation{nested-cuts-path-difference}
\end{align*}
These observations lead to the following bottom-up approach. We first
process $s_1$ as though it were a leaf, as described above.
% incrementing values on $\lca{s_1,x}$-to-root paths and decrementing
% on $s_1$-to-root paths for each edge $(s_1,x)$ incident to $s_1$,
% and then looking for small nested 2-cuts $D(t) - s_1$ by checking
% the minimum value on the $s_1$-to-root path, incrementing weights
% along good $D(t) - s_1$ min-cuts as they are found via the
% \refroutine{lazy-inc-cuts} data structure and propagating the
% changes to the link-cut trees.
For $i=1,\dots,\ell-1$, once we have certified that are no nested
$D(t) - s_i$ cuts of value $< (1 + \eps) \lambda$, we begin to process
$s_{i+1}$, continuing the aggregate values computed when processing
$s_i$ instead of starting over. For each edge $e = (s_{i+1},x)$
incident to $s_{i+1}$, we update the values along paths up the tree
per \refequation{nested-cuts-path-difference} as follows. For the
first term, $2\cutweight{s_{i+1},D(t)}$, we find the least common
ancestor $t = \lca{s_{i+1},x}$, and add $2 w_e$ to every vertex on the
$t$-to-root path. For the second term, $\cutweight{s_{i+1}}$, we add
$\weight{e}$ to every vertex on the $s_{i+1}$-to-root path. For the
third term, $2 \cutweight{s_{i+1},D(S_i)}$, if $x = s_j$ for some
$j < i$, then we subtract $2 \weight{e}$ from every vertex on the
$s_{i+1}$-to-root path. After processing the edges incident to
$s_{i+1}$, we repeatedly find the minimum value in the
$s_{i+1}$-to-root path so long as the minimum value is less than the
threshold $\epsmore \lambda$. Each time the minimum value is below the
threshold, we update weights along the edges of the corresponding cut
via \refroutine{inc-cut}, and propagate any increments returned by
\refroutine{inc-cut} (as before) to restore
\refequation{nested-cuts-path} before querying for the next minimum
value.

To process all the nested 2-cuts along the leaf-path, as with
incomparable 2-cuts along leaf-paths, the total amortized running time
is $O(\log n)$ times the number of edges incident to
$s_1,\dots,s_{\ell}$, plus $\bigO{\log^2 n}$ work for each cut output
and $\bigO{\log n}$ work for each edge increment returned by
\refroutine{inc-cut}.

\paragraph{All nested 2-cuts.} By the same induction step as for
incomparable 2-cuts, efficiently processing leaf-paths leads to a
procedure for efficiently processing the whole tree. The processing is
broken into phases. Each phase processes all the leaf-paths of the
current tree, and then contracts the leaf-paths into their
parents. Each leaf in the contracted tree had at least two children
previously, so the number of nodes has at least halved. After
$O(\log n)$ phases, we have processed the entire tree. Modulo
$\bigO{\log^2 n}$ work for each cut output and $\bigO{\log n}$ work
for each edge increment returned by \refroutine{inc-cut}, processing a
leaf path takes $\bigO{\log n}$ time for each edge incident to any
node on the path, and conversely each edge is incident to at most 2
leaf paths. Thus, excluding the time spent outputting cuts and
incrementing edge weights along the cuts, each phase takes
$\bigO{m \log n}$ to complete.

\begin{lemma}
  Let $T$ be a fixed and rooted spanning tree and $\lambda > 0$ a
  fixed target cut value. Employing the \refroutine{inc-cut} routing
  of the \refroutine{lazy-inc-cuts} data structure to manage edge
  weight increments, one can repeatedly find nested 2-cuts $C$ induced
  by $T$ of value $\cutweight{C} \leq (1 + \bigO(\eps)) \lambda$ and
  increment the corresponding edge weights (per the MWU framework)
  until certifying that there are no incomparable 2-cuts $C$ of value
  $\cutweight{C} \leq (1 + \eps) \lambda$, in
  $\bigO{m \log^2 n + K \log^2 n + I \log n}$ total running time,
  where $K$ is the number of nested 2-cuts of value
  $\leq (1 + O(\eps)) \lambda$ output and $I$ is the total number of
  weight increments returned by \refroutine{inc-cut}.
\end{lemma}

\section{Applying multiplicative weight updates along cuts}
\labelsection{lazy-cut-weights}

We address the remaining issue of implementing the
\refroutine{lazy-inc-cuts} data structure for a fixed and rooted
spanning tree $T$.  An interface for \refroutine{lazy-inc-cuts} was
given previously in \reffigure{lazy-inc-cuts-api} and we target the
bounds claimed by \reflemma{lazy-inc-cuts}. Recall that the MWU
framework takes a cut $C \in \cuts$, computes the smallest capacity
$\mincap = \argmin_{e \in C} \capacity{e}$ in the cut, and increases
the weight of each cut-edge $e \in C$ by a multiplicative factor of
$\expof{\eps \mincap / \capacity{e}}$ (see
\reffigure{held-karp-direct}). A subtle point is that the techniques
of \refsection{tree-cuts} identify approximate min-cuts without
explicitly listing the edges in the cut. A 1-cut $\cut{D(s)}$ is
simply identified by the root $s$ of the down-set $D(s)$, and likewise
2-cuts of the form $\cut{D(s) \cup D(t)}$ (when $s \incomp t$) and
$\cut{D(t) \setminus D(s)}$ (when $s < t$) can be described by the two
nodes $s$ and $t$. In particular, an approximately minimum cut $C$ is
identified without paying for the number of edges $\bigO{\sizeof{C}}$
in $C$ in the running time.  When it comes to updating the edge
weights, the natural approach of visiting each edge $e \in C$ would be
too slow, to say nothing of even identifying all the edges in $C$ in
time proportional to $\sizeof{C}$.

While the general task of incrementing\footnote{In the MWU framework
  the weights increase in a multiplicative fashion.  We use the term
  ``incrementing'' in place of ``updating'' since weights are
  increasing and also because it is convenient to think in terms of
  the logarithm of the weights, which do increase additively.}
weights along a cut appears difficult to execute both quickly and
exactly, we have already massaged the setting to be substantially
easier. First, we can afford to approximate the edge weights
$\weight{e}$ by a small multiplicative error. This means, for example,
that a cut edge $e \in C$ with very large capacity
$\capacity{e} \gg \mincap$ can to some extent be ignored. Second, we
are not incrementing weights along any cut, but just the 1-cuts and
2-cuts of a fixed rooted spanning tree $T$. We have already seen in
\refsection{tree-cuts} that restricting ourselves to 1-cuts and 2-cuts
allows us to (basically) apply dynamic programming to find small cuts,
and also allows us to use dynamic trees to efficiently update and scan
various values in the aggregate. Here too we will see that 1-cuts and
2-cuts are simple enough to be represented efficiently in standard
data structures. In the following, we use the term ``cuts'' liberally
as the set of edges with endpoints in each of two sets of vertices; in
particular, the two sets of vertices may not be disjoint.
\begin{lemma}
  \labellemma{canonical-cuts} Let $T$ be a fixed rooted spanning tree
  of an undirected graph $\defgraph$ with $\sizeof{\edges} = m$ edges
  and $\sizeof{\vertices} = n$ vertices. In $\bigO{m \log^2 n}$ time,
  one can construct a collection of nonempty cuts
  $\canonicalcuts \subseteq \subsetsof{\edges}$ such that
  \begin{properties}
  \item every edge $e \in \edges$ appears in at most $O(\log^2 n)$
    cuts $D \in \canonicalcuts$, and
  \item every 1-cut or 2-cut $C \in \cuts_T$ (described succinctly by
    at most 2 roots of subtrees) can be decomposed into the disjoint
    union
    \begin{math}
      C = D_1 \dunion \cdots \dunion D_{\ell}
    \end{math}
    of $\ell = O(\log^2 n)$ cuts
    $D_1,\dots,D_{\ell} \in \canonicalcuts$ in $O(\log^2 n)$ time.
  \end{properties}
\end{lemma}
By building the collection $D \in \canonicalcuts$ once for a tree $T$,
\reflemma{canonical-cuts} reduces the problem of incrementing along
any 1-cut or 2-cut $C \in \cuts_T$ to incrementing along a
``canonical'' cut $D \in \canonicalcuts$ known \emph{a priori}. This
is important because multiplicative weight updates can be applied to a
\emph{static} set relatively efficiently by known techniques
\citep{y-14,cq-17}. It is also important that cuts in $\canonicalcuts$
are sparse, in the sense that
$\sum_{D \in \canonicalcuts} \sizeof{D} = \bigO{m \log^2 n}$, as this
sum factors directly into the running time guarantees. We first prove
\reflemma{canonical-cuts} in \refsection{canonical-cuts}, and then in
\refsection{lazy-incs} we show how to combine amortized data
structures for each $D \in \canonicalcuts$ to increment along any
1-cut or 2-cut $C \in \cuts_T$.

\subsection{Canonical cuts}
\labelsection{canonical-cuts}

\begin{wrapfigure}{r}{.28\textwidth}

  \vspace{-56pt}

  \centerline{\includegraphics[width=.2\textwidth]{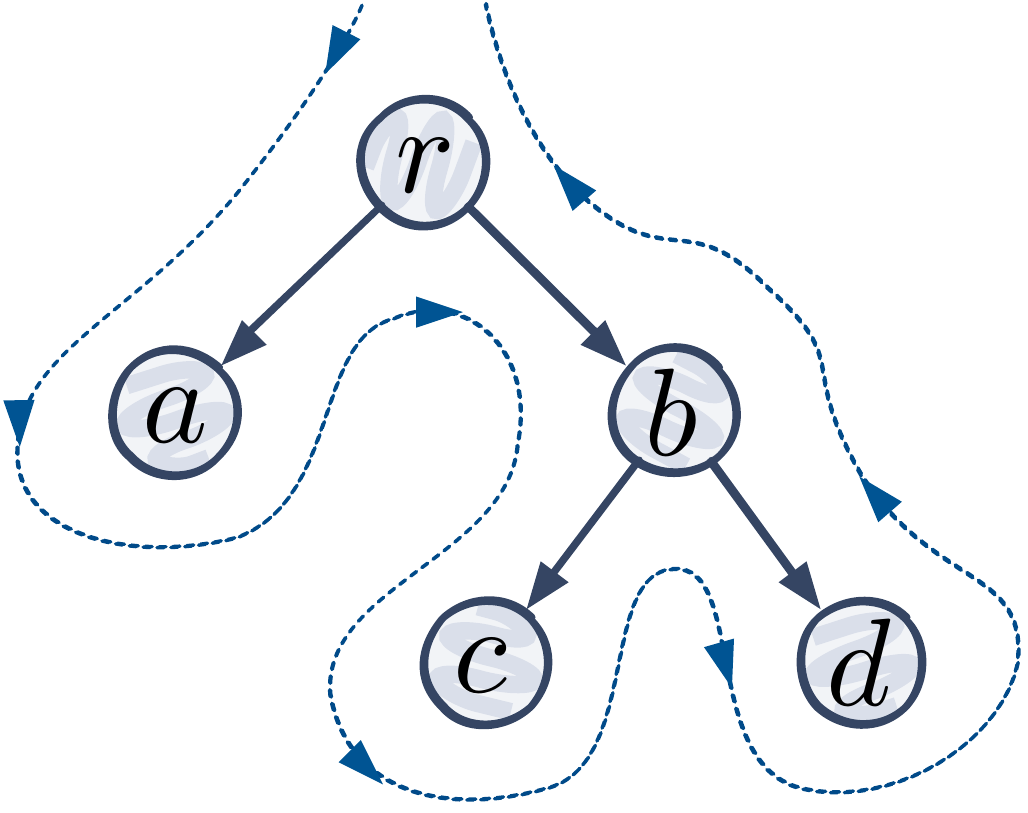}}

  \footnotesize An Euler tour inducing the ordering
  \begin{math}
    r^- < a^- < a^+ < b^- < c^- < c^+ < d^- < d^+ < b^+ < r^+.
  \end{math}

  \vspace{-3ex}

\end{wrapfigure}
\textbf{Euler tours and orderings.}  Let $\tree$ be a fixed and rooted
tree on $n$ vertices $\vertices$, and fix an Euler tour on a
bidirected copy of $\tree$ that replaces each edge of $\tree$ with
arcs in both directions, starting from the root. For each vertex $v$,
we create two symbols: $v^-$ means we enter the subtree $\subtree{v}$
rooted at $v$, and $v^+$ means we leave the subtree $v$. Let
\begin{math}
  \vertices^{\pm} = \setof{v^-, v^+ \where v \in \vertices{\tree}}
\end{math}
denote the whole collection of these $2n$ symbols. The Euler tour
enters and leaves each subtree exactly once in a fixed order. Tracing
the Euler tree induces a unique total ordering on $\vertices^\pm$ (see
the picture on the right).

This ordering has a couple of interesting properties.  For every
vertex $v \in \vertices(\tree)$, we have $v^- < v^+$.  Letting $r$
denote the root of $\tree$, $r^-$ is the first element in the ordering
and $r^+$ is the last element in the ordering. More generally, for a
vertex $v$ and a vertex $w\in D(v)$ in the subtree rooted at $v$, we
have $v^- \leq w^- < w^+ \leq v^+$, with all inequalities strict if
$v \neq w$. That is, the ranges
$\setof{[v^-,v^+] : v \in \vertices(\tree)}$ between entering and
leaving a subtree form a laminar set.

The Euler order on the vertices endows a sort of geometry to the
edges. Each edge $e = (u,v)$ (with $u^- < v^-$) can be thought of as
an interval $[u^-, v^-]$. A downset $D(s)$ cuts $e$ iff
\begin{math}
  u^- \leq s^- \leq v^- \leq s^+
\end{math}
or
\begin{math}
  s^- \leq u^- \leq s^+ \leq v^-
\end{math}
(see \reffigure{euler-order-tree-cuts}); that is, iff
\begin{math}
  \sizeof{\setof{u^-,v^-} \cap [s^-,s^+]} = 1.
\end{math}
Alternatively, $D(s)$ cuts $e$ iff
\begin{math}
  \sizeof{[u^-,v^-] \cap \setof{s^{-},s^+}} = 1.
\end{math}
These observations suggest that this is not a problem about graphs,
but about intervals, and perhaps a problem suited for range trees.

\begin{figure}
  \hfill%
  \hfill%
  \includegraphics[width=.21\textwidth]{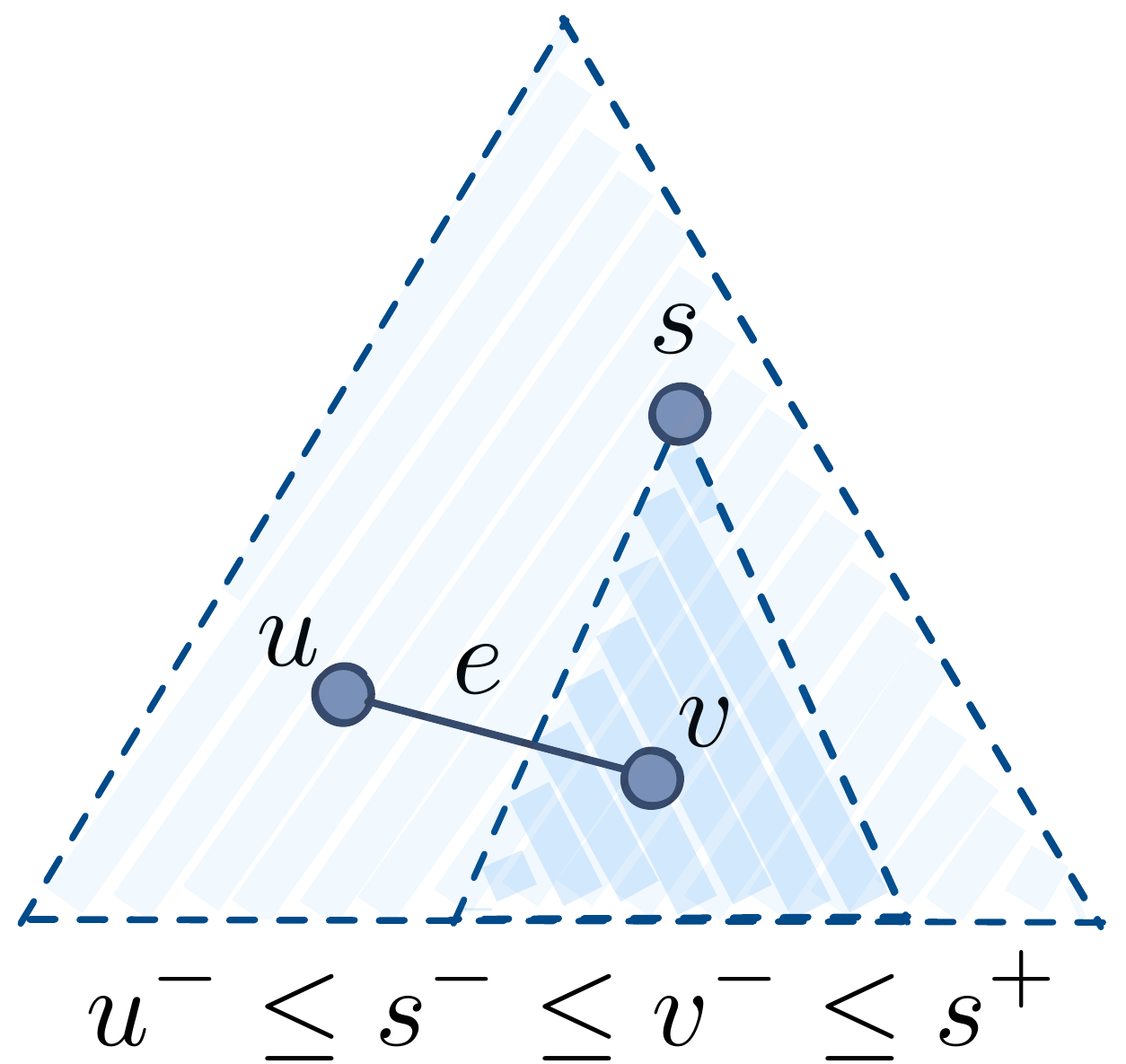}%
  \hfill%
  \includegraphics[width=.21\textwidth]{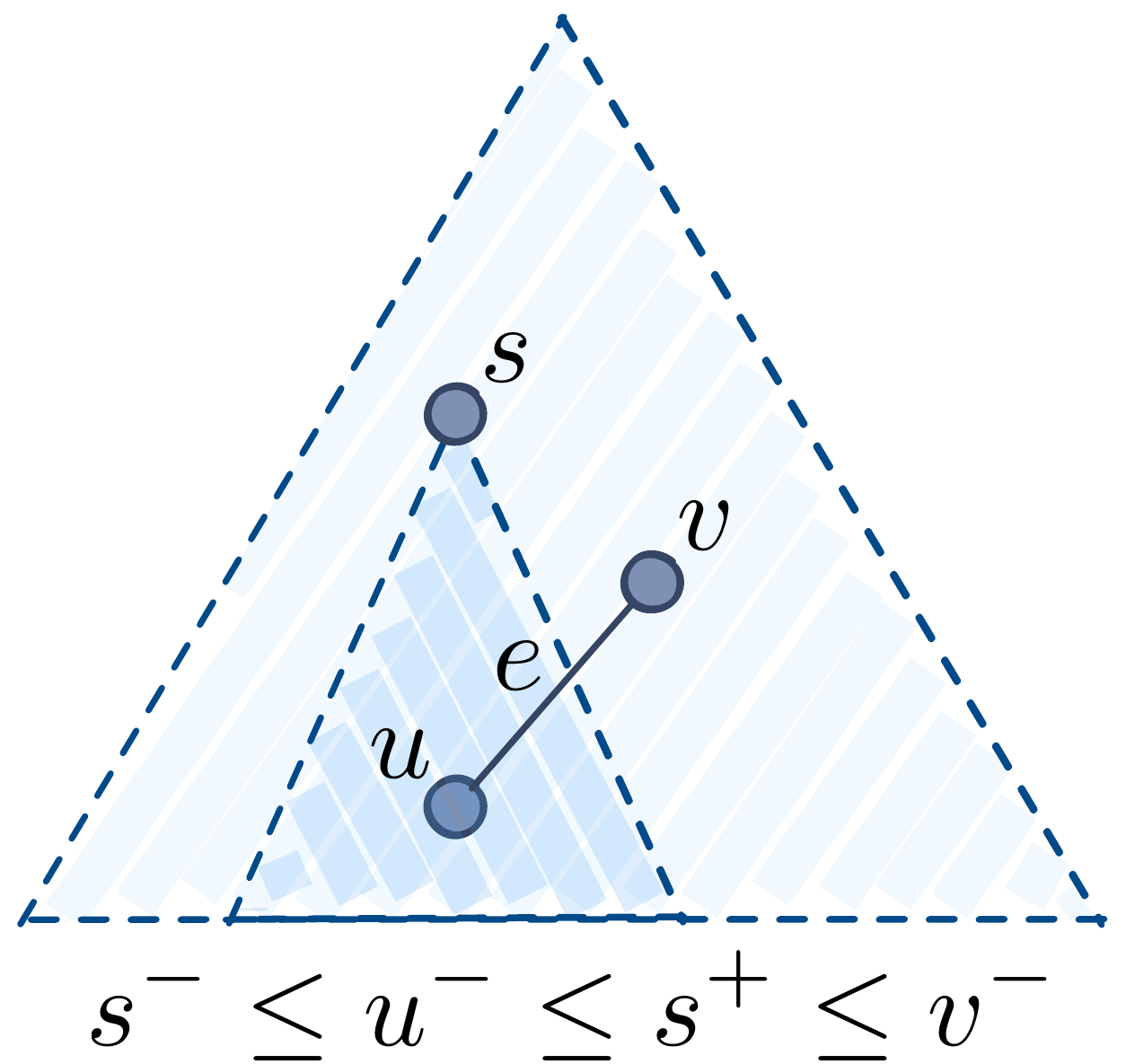}%
  \hfill%
  \hfill%
  \hfill%

  \caption{An edge cut $e$ by a subtree $\subtree{s}$, and two
    possible inequalities in the Euler order.
    \labelfigure{euler-order-tree-cuts}}
\end{figure}

\paragraph{Range trees on the Euler order.}  Let $R$ be a balanced
range tree with $\vertices^{\pm}$ at its leaves. As a balanced tree
with $2n$ leaves, $R$ has $\bigO{n}$ nodes and height $O(\log
n)$. Each range-node $a \in \vertices{R}$ induces an interval $I_a$ on
$\vertices^{\pm}$, consisting of all the elements of $\vertices^{\pm}$
at the leaves of the subtree $R_a$ rooted at $a$. We call $I_a$ a
\emph{canonical interval} of $\vertices^{\pm}$ (induced by $a$), and
let $\intervals = \setof{I_a: a \in \vertices(R)}$ denote the
collection of all $O(n)$ canonical intervals.

\begin{wrapfigure}{o}{.2\textwidth}
  \vspace{-6ex}

  \smallskip

  \includegraphics[width=.2\textwidth]{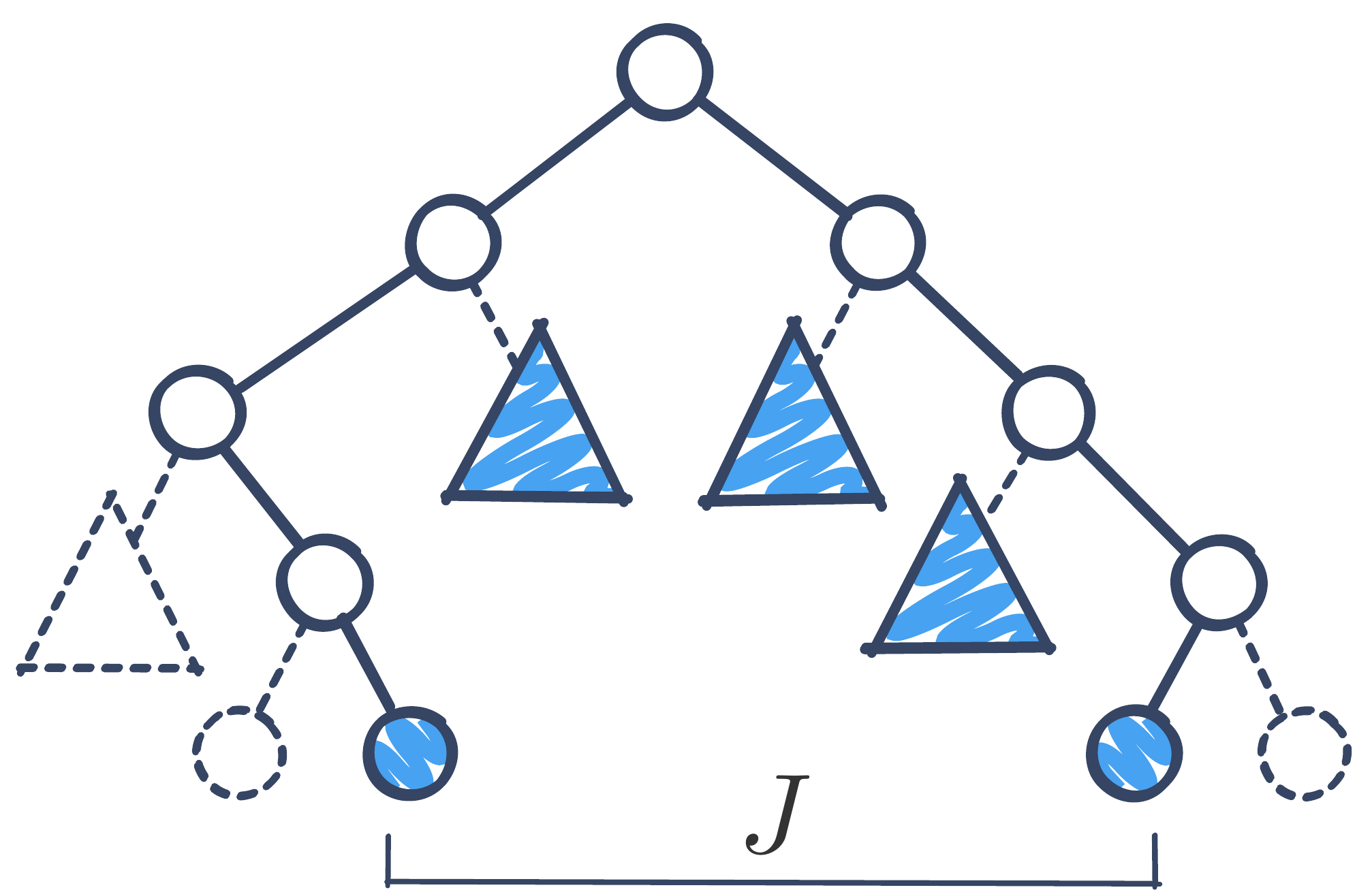}

  \smallskip

  \vspace{-3ex}
\end{wrapfigure}
Each element in $\vertices^{\pm}$ appears in $O(\log n)$ canonical
intervals because the height of $R$ is $O(\log n)$. Moreover, every
interval $J$ on $\vertices^{\pm}$ decomposes into the disjoint union
of $O(\log n)$ canonical intervals. The decomposition can be obtained
in $O(\log n)$ time by tracing the two paths from the root of $R$ to
the endpoints of $J$ and taking the canonical intervals corresponding
to the $O(\log n)$ maximal subtrees between the paths (see the picture
on the right). Similarly, the union of a constant number of intervals
or the complement of the union of a constant number of intervals
decomposes to $O(\log n)$ canonical intervals in $O(\log n)$ time.

The canonical intervals $\intervals$ relate to the 1-cuts and 2-cuts
$\cuts_T$ induced by $T$ as follows. For any pair of disjoint
canonical intervals $I_1, I_2 \in \intervals$, let
\begin{math}
  \cut{I_1,I_2} %
  = %
  \setof{(u,v) \in \edges \suchthat u \in I_1, v \in I_2}
\end{math}
be the set of edges with one endpoint in $I_1$ and the other in
$I_2$. We call $\cut{I_1,I_2}$ a \emph{canonical cut} induced by $I_1$
and $I_2$.  We claim that any 1-cut or 2-cut $C \in \cuts_T$
decomposes into the disjoint union of $O(\log^2 n)$ canonical cuts.

\begin{wrapfigure}{o}{.18\textwidth}
  \vspace{-24pt}

  \includegraphics[width=.18\textwidth]{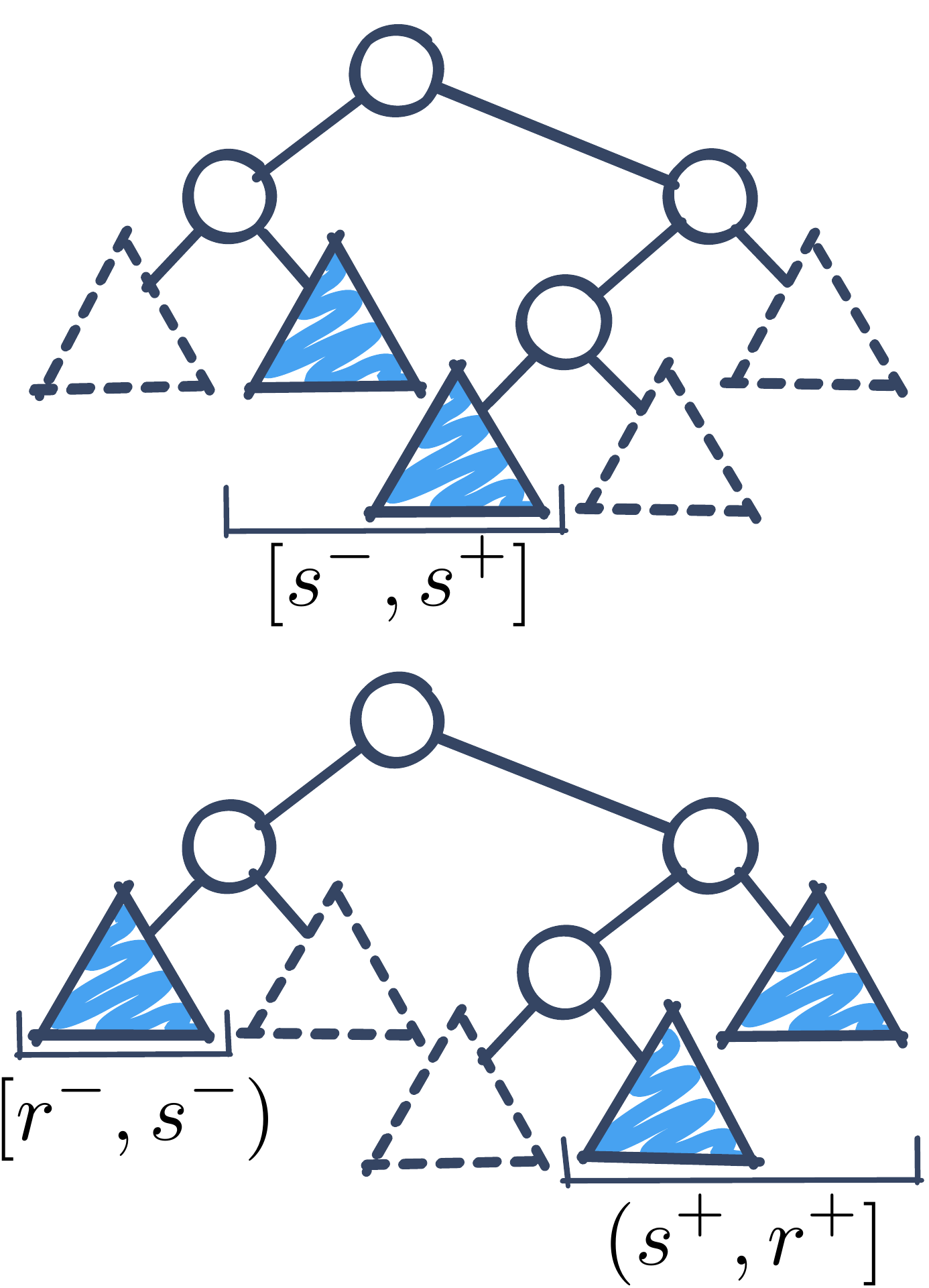}

  \vspace{-48pt}
\end{wrapfigure}
For starters, let $C = \cut{D(s)}$ be the 1-cut induced by the downset
of a vertex $s$. The 1-cut $\cut{D(s)}$ consists of all edges $(u,v)$
with one (tagged) endpoint $u^-$ in the interval $[s^-,s^+]$ and the
other endpoint $v^-$ outside $[s^-,s^+]$. The interval $[s^-,s^+]$
decomposes to the disjoint union
\begin{math}
  [s^-, s^+] = \Dunion_{I_1 \in \intervals_1} I_1
\end{math}
of $O(\log n)$ canonical intervals
$\intervals_1 \subseteq \intervals$, and the complement
$[r^-, s^-) \cup (s^+,r^+]$ also decomposes to the disjoint union
\begin{math}
  [r^-, s^-) \cup (s^+,r^+] %
  = %
  \Dunion_{I_2 \in \intervals_2} I_2
\end{math}
of $O(\log n)$ canonical intervals $\intervals_2 \subseteq \intervals$
(see the picture on the right). Together, the disjoint union
\begin{math}
  \cut{D(s)} = \Dunion_{I_1 \in \intervals_1, I_2 \in \intervals_2}
  \cut{I_1,I_2}
\end{math}
of canonical cuts over the cross product
$(I_1,I_2) \in \intervals_1 \times \intervals_2$ decomposes the 1-cut
$\cut{D(s)}$ into
$\sizeof{\intervals_1} \times \sizeof{\intervals_2} = O(\log^2 n)$
canonical cuts. Moreover, both decompositions $\intervals_1$ and
$\intervals_2$ can be obtained in $O(\log n)$ time.

Any 2-cut decomposes similarly. If $s \incomp t$ are incomparable,
with (say) $s^- < s^+ < t^- < t^+$, then the incomparable 2-cut
$C = \cut{D(s) \cup D(t)}$ is the set of edges with one endpoint in
$[s^-, s^+] \cup [t^-,t^+]$ and other endpoint in the complement,
$[r^-, s^-) \cup (s^+, t^-) \cup (t^+,r^+]$. Both of these sets are
the disjoint union of two or three intervals, and they each decompose
into $O(\log n)$ canonical intervals. The cross-product of the two
decompositions decomposes $\cut{D(s) \cup D(t)}$ into
$\bigO{\log^2 n}$ canonical cuts.

If $s< t$ are two comparable vertices, then $t^- < s^- < s^+ < t^+$
and the nested 2-cut $\cut{D(t) \setminus D(s)}$ is the set of edges
with one endpoint in $[t^-,s^-) \cup (s^+, t^+]$ and the other
endpoint in the complement $[r^-,t^-) \cup [s^-,s^+] \cup [t^+,r^+]$
and the cross-product breaks $\cut{D(t) \setminus D(s)}$ down into
$\bigO{\log^2 n}$ disjoint canonical cuts.

Thus, any 1-cut or 2-cut $C \in \cuts_T$ breaks down into to the
disjoint union of $O(\log^2 n)$ canonical cuts. An edge $e = (u,v)$
appears in a canonical cut $\cut{I_1,I_2}$ iff $u \in I_1$ and
$v \in I_2$. As either end point $u$ or $v$ appears in $O(\log n)$
canonical intervals, $e$ appears in at most $O(\log^2 n)$ canonical
cuts.  In turn, there are at most $O(m \log^2 n)$ nonempty canonical
cuts. The nonempty canonical cuts are easily constructed in
$O(m \log^2 n)$ time by adding each edge to the $\bigO{\log^2 n}$
canonical cuts containing it. Taking $\canonicalcuts$ to be this set
of nonempty canonical cuts gives \reflemma{canonical-cuts}.

\subsection{Lazy weight increments}
\labelsection{lazy-incs} %

\reflemma{canonical-cuts} identifies a relatively small set of
canonical cuts $\canonicalcuts$ such that any 1-cut or 2-cut
$C \in \cuts_T$ is the disjoint union of $\bigO{\log^2 n}$ canonical
cuts $D_1,\dots,D_{\bigO{\log^2 n}} \in \canonicalcuts$. While we can
now at least gather an implicit list of all the edges of $C$ in
$\bigO{\log^2 n}$ time, it still appears necessary to visit every edge
$e \in C$ to increment its weight. The challenge is particularly
tricky because the multiplicative weight updates are not
uniform. However, for a fixed and static set, recent techniques by
\citet{y-14} and \citet{cq-17} show how to approximately apply a
multiplicative weight update to the entire set in polylogarithmic
amortized time. In this section, we apply these techniques to each
canonical cut $D \in \canonicalcuts$ and show how to combine their
outputs carefully to meet the claimed bounds of
\reflemma{lazy-inc-cuts}.

For every canonical cut $D \in \canonicalcuts$, we employ the
\algo{lazy-inc} data structure of \citet{cq-17}. The
\refroutine{lazy-incs} data structure gives a clean interface to
similar ideas in \citet{y-14}, and we briefly describe a restricted
version of the data structure in \citep{cq-17} per the needs of this
paper\footnote{What we refer to as \refroutine{lazy-incs} is most
  similar to \refroutine{lazy-incs}{0} in \citet{cq-17}.}. The
\algo{lazy-incs} data structure is an amortized data structure that
approximates a set of counters increasing concurrently at different
rates. For a fixed instance of \refroutine{lazy-incs} for a particular
canonical cut $D \in \canonicalcuts$, we will have one counter for
each edge tracking the ``additive part''
$v_e = \lnof{\weight{e}} / \eps$ for each edge $e \in D$. The rate of
each counter $e$ is stored as $\rate{e}$, and in this setting is
proportional to the inverse of the capacity $\capacity{e}$. The
primary routine is \refroutine{inc}{\varrate}, which simulates a
fractional increase on each counter proportional to a single increment
at the rate $\varrate$.  That is, each counter $v_e$ increases by
$\rate{e} / \varrate$.  The argument $\varrate$ will always be taken
to be proportional to the inverse of the minimum capacity of any edge
in the cut, $\mincap = \min_{e \in C} \capacity{e}$.

The output of \refroutine{inc}{\varrate} is a list of increments
$\setof{(e,\delta_e)}$ over some of the edges in $D$. The routine does
not return an increment $(e,\delta_e)$ unless $\delta_e$ is
substantially large. This allows us to charge off the work required to
propagate the increment $(e,\delta_e)$ to the rest of the algorithm to
the maximum weight of $e$.  In exchange for reducing the number of
increments, the sum of returned increments for a fixed edge $e$
underestimates $v_e$ by a small additive factor.  An underestimate for
$v_e = \lnof{\weight{e}} / \eps$ within an additive factor of $O(1)$
translates to an underestimate for $\weight{e} = \expof{\eps v_e}$
within a $(1 + O(\eps))$-multiplicative factor, as desired.

At a high level, \refroutine{lazy-incs} buckets the counters by
rounding up each rate to the nearest power of 2. For each power of 2,
an auxiliary counter is maintained with rate set to this power (more
or less) exactly and efficiently.  Counters with rates that are powers
of 2 can be maintained in constant amortized time for the same reason
that a binary number can be incremented in constant amortized
time. When an auxiliary counter increases to the next whole integer,
the tracked counters in the corresponding bucket are increased
proportionately. Up to accounting details, this lazy scheme tracks the
increments of each counter to within a constant additive factor at all
times. A more general implementation of \refroutine{lazy-incs} and a
detailed proof of the following theorem are given in \citep{cq-17}.

\begin{lemma}[{\citealp{cq-17}}]
  \labellemma{lazy-incs} Let $e_1,\dots,e_k$ be $k$ counters with
  rates $\rate{e_1} \geq \rate{e_2} \geq \cdots \geq \rate{e_k}$ in
  sorted order.
  \begin{mathresults}
  \item An instance of \refroutine{lazy-incs} can be initialized in
    $O(k)$ time.
  \item Each \refroutine{inc} runs in $\bigO{1}$ amortized time plus
    $O(1)$ for each increment returned.
  \item \refroutine{flush} runs in $O(k)$ time.
  \item For each counter $e_i$, let $v_i$ be the true value of the
    counter and let $\tilde{v}_i$ be the sum of increments for counter
    $i$ in the return values of \refroutine{inc} and
    \refroutine{flush}.
    \begin{mathproperties}
    \item After each call to \refroutine{inc}, we have
      $\tilde{v}_i \leq v_i \leq \tilde{v}_i + O(1)$ for each counter
      $e_i$.
    \item After each call to \refroutine{flush}, we have
      $\tilde{v}_i = v_i$ for each counter $e_i$.
    \end{mathproperties}
  \end{mathresults}
\end{lemma}

We note that \refroutine{flush} is not implemented in \citep{cq-17},
but is easy enough to execute by just reading off all the residual
increments in the data structure and resetting these values to 0.

The application of \refroutine{lazy-incs} to our problem bears some
resemblance to its application to packing intervals in \citep{cq-17},
where the weights are also structured by range trees.  We first invoke
\reflemma{canonical-cuts} to generate the family of canonical cuts
$\canonicalcuts$. For each canonical cut $D \in \canonicalcuts$, by
\reflemma{lazy-incs}, we instantiate an instance of
\refroutine{lazy-incs} where each edge $e \in D$ has rate
\begin{math}
  \rate{e} = \log^2 n / \capacity{e}.
\end{math}
(\reflemma{lazy-incs} takes as an input a collection of counters
sorted by rate. Rather than sort each canonical cut individually, we
can save a $\logof{n}$-factor by a slight modification to
\reflemma{canonical-cuts}. If the edges are sorted before applying
\reflemma{canonical-cuts}, then each canonical cut
$D \in \canonicalcuts$ will naturally be in sorted order.) We also
compute, for each canonical cut $D \in \canonicalcuts$, the minimum
capacity $\mincap_D = \min_{e \in D} \capacity{e}$ in the cut. These
preprocessing steps take $\bigO{m \log^2 n}$ time total.

When incrementing weights along a cut $C \in \cuts_T$, we update the
weights as follows. \reflemma{canonical-cuts} divides $C$ into the
union of $O(\log^2n)$ disjoint canonical cuts
$\canonicalcuts' \subseteq \canonicalcuts$. We find the minimum
capacity $\mincap = \min_{e \in C} \capacity{e}$ by taking the minimum
of the precomputed minimum capacities of each canonical cut
$D \in \canonicalcuts'$, i.e.,
$\mincap = \min_{D \in \canonicalcuts'} \mincap_{D}$. For each
canonical set $D \in \canonicalcuts'$, we call \refroutine{inc}{\log^2
  n / \mincap} on the corresponding \refroutine{lazy-incs}
instance. For every increment $(e,\delta)$ returned in
\refroutine{lazy-incs}, where $e \in C$ and $\delta > 0$ represents a
increment to $v_e = \lnof{\weight{e}}/\eps$, we increase $v_e$ by an
additive factor of $\delta / \log^2 n$, and multiply $\weight{e}$ by
$\expof{\eps \delta / \log^2 n}$ accordingly.

Fix an edge $e \in \edges$. The edge $e$ is a member of
$\bigO{\log^2 n}$ canonical cuts in $\canonicalcuts$, and receives its
weight increments from $O(\log^2 n)$ instances of
\refroutine{lazy-incs}. The slight errors from each instance of
\refroutine{lazy-incs} accumulate additively (w/r/t
$v_e = \lnof{\weight{e}}/\eps$). Although each instance of
\refroutine{lazy-incs} promises only a constant additive error in
\reflemma{lazy-incs}, we scaled up the rate of $e$ from
$1 / \capacity{e}$ to $\log^2 n / \capacity{e}$ and divide the
returned weight increments by a factor of $\log^2 n$. The scaling
reduces the error from each instance of \refroutine{lazy-incs} from an
additive factor of $O(1)$ to $O(1 / \log^2 n)$. Consequently, the sum
of all $O(\log^2 n)$ instances of \refroutine{lazy-incs}
underestimates $v_e$ to an $O(1)$ additive error w/r/t $v_e$, and
underestimates $\weight{e}$ by at most a
$(1 + O(\eps))$-multiplicative factor, as desired.

Increasing the sensitivity of each \refroutine{lazy-incs} data
structure by a factor of $\log^2 n$ means every increment $(e,\delta)$
returned by \refroutine{inc} may increase $v_e$ by as little as
$1 / \log^2 n$. An additive increase in $1/\log^2 n$ corresponds to a
multiplicative increase of $\expof{\eps / \log^2 n}$ in $\weight{e}$,
hence property \refitem{lazy-inc-cuts-min-increment} in
\reflemma{lazy-inc-cuts}.

Together, the construction of the canonical cuts $\canonicalcuts$ and
the $O(m \log^2 n)$ carefully calibrated instances of
\refroutine{lazy-inc}, one for each canonical cut
$D \in \canonicalcuts$, give \reflemma{lazy-inc-cuts}.

\section{Putting it all together: Proof of \reftheorem{main}}
\labelsection{putting-together} In this section we combine the
ingredients discussed so far and outline the proof of
\reftheorem{main}.  Let $(\graph,\capacity)$ be an \mtsp instance and
$\eps' > 0$ be the error parameter.  We assume without loss of
generality that that $\eps'$ is sufficiently small, and in particular
that $\eps' < 1/2$. We will also assume that $\eps' > 1/n^2$ for
otherwise one could use an exact algorithm and achieve the desired
time bound. This implies that $\log \frac{1}{\eps'} = O(\log n)$.

Consider the algorithm in \reffigure{held-karp-lazy-inc-cuts} which
takes $\eps$ as a parameter. We will choose $\eps = \rho \eps'$ for
some sufficiently small $\rho$. It thus suffices to argue that the
algorithm outputs a $(1 + O(\eps))$-approximation with high
probability.

\paragraph{High-level MWU analysis:} At a high-level, our algorithm is
a standard width-independent MWU algorithm for pure packing problems
with an $\alpha$-approximate oracle, for $\alpha = (1+O(\eps))$. Our
implementation follows the ``time''-based algorithm of \citet{cjv-15}.
An exact oracle in this setting repeatedly solves the global minimum
cut problem w/r/t the edge weights $\weight$. To implement an
$\alpha$-approximate oracle, every cut output needs to be an
$\alpha$-approximate minimum cut.

To argue that we are indeed implementing an $\alpha$-approximate
minimum cut oracle, let us identify the two points at which we deviate
from Karger's minimum cut algorithm \cite{k-00}, which would otherwise
be an exact oracle that succeeds with high probability. In both
Karger's algorithm and our partially dynamic extension, we sample
enough spanning trees from an approximately maximum tree packing to
contain all $(1 + \bigO{\eps})$-approximate minimum cuts as a 1-cut or
2-cut with probability $1 - \poly{1/n} = 1 - \poly{\eps/n}$ (see
\reftheorem{karger-tree-packing}). The sampled tree packings are the
only randomized component of the algorithm, and we will argue that all
the sampled tree packings succeed with high probability later after
establishing a basic correctness in the event that all the sampled
packings succeed.

Karger's algorithm searches all the spanning trees for the minimum
1-cut or 2-cut. We retrace the same search, except we output any
approximate minimum 1-cut or 2-cut found in the search. More
precisely, we maintain a target value $\lambda$ with the invariant
that there are no cuts of value $< \lambda$, and output any 1-cut or
2-cut with weight $\leq (1 + \bigO{\eps})\lambda$ at that moment (see
\reflemma{cut-search}). Since $\lambda$ is no more than the true
minimum cut w/r/t $\weight$ at any point, any cut with weight
$\leq (1 + \bigO{\eps})\lambda$ is a $(1+\bigO{\eps})$-approximation
to the current minimum cut.

The second point of departure is that we do not work with the ``true''
weights implied by the framework, but a close approximation. By
\reflemma{lazy-inc-cuts}, the approximated weights underestimate the
true weights by at most a $(1 + \bigO{\eps})$-multiplicative
factor. In turn, we may underestimate the value of a cut by at most a
$(1 + \bigO{\eps})$-multiplicative factor, but crucially any 1-cut or
2-cut will still have value $\leq (1 + \bigO{\eps})\lambda$, for a
slightly larger constant hidden in the $\bigO{\eps}$.

This firmly establishes that the proposed algorithm in fact implements
an $\alpha$-approximate oracle for $\alpha = 1 +\bigO{\eps}$ (with
high probability). The MWU framework then guarantees that the final
packing of cuts output by the algorithm approximately satisfies all
the constraints with approximately optimal value, as desired.

\paragraph{Probability of failure.} We now consider the probability of
the algorithm failing. Randomization enters the algorithm for two
reasons. Initially, we invoke Karger's randomized minimum cut
algorithm \cite{k-00} to compute the value of the minimum cut w/r/t
the initial weights $\weight = 1/c$, in order to set the first value
of $\lambda$.  Second, at the beginning of each epoch, we randomly
sample a subset of an approximately maximum tree packing hoping that
the sampled trees contain every $(1 + O(\eps))$-approximate minimum
cut as a 1-cut or 2-cut in one of the sampled
trees\footnote{Interestingly, the only randomization in Karger's
  algorithm also stems from randomly sampling an approximately maximum
  tree packing.}.

\citet{k-00} showed that the minimum cut algorithm fails with
probability at most $\poly{1/n}$. He also showed (see
\reftheorem{karger-tree-packing}) that each random sample of
$\bigO{\logof{n/\eps}} = \bigO{\log n}$ spanning trees of an
approximately maximum tree packing (conducted at the beginning of an
epoch) fails to capture all $(1+\eps)$-approximate minimum cuts with
probability at most $\poly{1 / n} = \poly{\eps / n}$. As discussed in
\refsection{tree-packings}, there are a total of
$\bigO{\log n / \eps^2}$ epochs over the course of the algorithm. By
the union bound, the probability of \emph{any} random sample of
spanning trees failing is
\begin{math}
  \bigO{\log n / \eps^2} \cdot \poly{\eps / n} = \poly{\eps / n}.
\end{math}
Taking a union bound again with the chance of the initial call to
Karger's minimum cut algorithm failing, the probability of any
randomized component of the algorithm failing is at most $\poly{1/n}$,
as desired.

\paragraph{Running time.} It remains to bound the running time of the
algorithm. The analysis is not entirely straightforward, as some
operations can be bounded directly, while others are charged against
various upper bounds given by the MWU framework.

The algorithm initializes the edge weights in $\bigO{m}$ time, and
invokes Karger's minimum cut algorithm \cite{k-00} once which runs in
$\bigO{m \log^3 n}$ time.  As established in
\refsection{tree-packings}, the remaining algorithm is divided into
$\bigO{\logof{n} / \eps^2}$ epochs. Each epoch invokes
\reftheorem{karger-tree-packing} once to sample
$\bigO{\logof{n/\eps}} = \bigO{\log n}$ spanning trees from an
approximate tree packing in $\bigO{m \log^3 n}$ time. We then invoke
\reflemma{cut-search} for each spanning tree in the sample. Over the
course of $\bigO{\logof{n} / \eps^2}$ epochs, each with
$\bigO{\logof{n}}$ spanning trees, we invoke \reflemma{cut-search} at
total of $\bigO{\log^2(n) / \eps^2}$ times.

Suppose we process a tree $T$, outputting $K$ approximately minimum
cuts and making $I$ full edge weight increments. By
\reflemma{cut-search}, the total time to process $T$ is
$\bigO{m \log^2 n + K \log^2 n + I \log n}$. The first term,
$\bigO{m \log^2 n}$, comes from following the tree-processing
subroutine of \citet{k-00}. Over a total $\bigO{\log^2(n) / \eps^2}$
trees, the total time spent mimicking Karger's subroutine is
$\bigO{m \log^4(n) / \eps^2}$.

The remaining terms of \reflemma{cut-search}, depending on $K$ and
$I$, are introduced by the MWU framework and can be amortized against
standard properties of the MWU framework. The quantity $K$ represents
the number of approximate minimum cuts output while processing a
single tree $T$, and the sum of all $K$ across all trees is the total
number of approximate minimum cuts output. Each such cut corresponds
to a single iteration, and the total number of iterations in the MWU
framework is $\bigO{m \logof{n} / \eps^2}$. Thus, the sum of the
second term $\bigO{K \log^2 n}$ over all invocations of
\reflemma{cut-search} is $\bigO{m \log^3 (n) / \eps^2}$.

The third term of \reflemma{cut-search}, $\bigO{I \log n}$ depends on
the number of increments $I$ returned by the \refroutine{inc-cut}
subroutine of the \refroutine{lazy-inc-cuts} data structure. Each
increment $(e,\delta)$ returned by \refroutine{inc-cut} increases
$\weight{e}$ by at least a
$(1 + \Omega(\eps / \log^2 n))$-multiplicative factor (i.e.,
$\delta \geq \Omega(\eps \weight{e} / \log^2 n)$).  The MWU framework
shows that a single edge can increase by a
$(1 + \Omega(\eps / \log n))$ factor at most
$\bigO{\log^3(n) / \eps^2}$ times (see \refsection{mwu}). It follows
that the total number of edge increments, over all edges, returned by
\refroutine{lazy-inc-cuts} is $\bigO{m \log^3(n) / \eps^2}$. Thus, the
sum of the quantities $\bigO{I \log n}$ over all invocations of
\reflemma{cut-search} is $\bigO{m \log^4(n) / \eps^2}$.

All told, the algorithm runs in
\begin{math}
  \bigO{m \log^4 (n) / \eps^2} = \apxO{m / \eps^2}
\end{math}
time total.

%%%%%%%%%%%%%%%%%%%%%%%%%%%%%%%%

\paragraph{Acknowledgements.} We thank Neal Young for suggesting that
\mtsp may be amenable to the techniques from \citep{cq-17}.

\bibliographystyle{abbrvnat} \bibliography{held-karp} %

\appendix

\section{Computing the primal solution}
Our MWU algorithm computes a $(1-\eps)$-approximate solution to
\ecssd. Recall that the algorithm maintains a weight $w(e)$ for each
edge $e$. These weights evolve with time. Using a standard argument we
show how we can recover a $(1+\eps)$-approximation to the LP \ecss.

Let $\opt$ be the optimum solution value of $\ecssd$ and $\ecss$.  Our
algorithm computes an implicit packing in the LP $\ecssd$ of value
$(1-\eps)\opt$ in near-linear time. It does this via the MWU method
which works in several iterations.  In each iteration $i$ the
algorithm computes a global mincut $C_i$ with respects to the edge
weights $w_i(e)$ where $w_i(e)$ is the weight of edge $e$ in iteration
$i$.  Each iteration corresponds to solving a relaxation of the
original LP by collapsing the constraints into a single constraint via
the weights in iteration $i$. The value of the solution to this
relaxation in iteration $i$ is $(\sum_e w_i(e) c(e))/ w_i(C_i)$.
Since this is a relaxation to the original LP, the value of this
solution is at least $\opt$.  Over the entire algorithm we have a time
of one unit, and the integral of all solutions we compute is at least
$\opt$. Let $z$ be the averaged solution at the end of the
algorithm. Via the MWU analysis $z$ violates each constraint by at
most a $(1+\eps)$-factor. Thus by scaling down $z$ by $(1+\eps)$, we
ensure feasibility of constraints and obtain a feasible solution to
\ecssd of value at least $\frac{1}{1+\eps}\opt$.

We claim (and this is standard) that there is an iteration $i$ such
that the value of the solution
$\sum_e w_i(e) c(e)/ w_i(C_i) \le (1+\eps) \opt$. If not, the value of
$z$ would be larger than $(1+\eps) \opt$ and we would, after scaling
down by a factor of $(1+\eps)$, obtain a feasible solution to \ecssd
of value more than $\opt$, a contradiction.  Given an iteration $i$
where $\sum_e w_i(e) c(e)/ w_i(C_i) \le (1+\eps) \opt$ we can obtain a
feasible solution to \ecss as follows.  Simply set
$y(e) = w_i(e)/w_i(C_i)$ for each $e$; we claim that this is a
feasible solution whose cost is at most $(1+\eps)\opt$. To see that
$y$ is feasible, consider any cut $C_j$. The capacity of the cut is
$\sum_{e \in C_j} y_e = w_i(C_j)/w_i(C_i) \ge 1$ since $C_i$ is the
cheapest cut with respect to the weights $w_i(e)$. The cost of the
solution $y$ is
$\sum_e y(e)c(e) = \sum_e w_i (e) c(e) / w_i(C_i) \le (1+\eps)\opt$ by
choice of $i$.

The solution $y$ can be output in nearly linear time.  Our algorithm
maintains the weights $w_i(e)$ implicitly via data
structures. However, during iteration $i$ we know the value of the
mincut $w_i(C_i)$ to within a $(1+O(\eps))$ factor as well as the sum
$\sum_e w_i(e) c(e)$. In particular, we can keep track of the
iteration $i^*$ that minimizes this ratio; this iteration $i^*$ will
satisfy the desired condition,
$\sum_e w_{i^*}(e) c(e)/ w_{i^*}(C_{i^*}) \leq (1+O(\eps)) \opt$.  The
actual set of weights $w_{i^*}(e)$ needed to create the solution
$y(e)$ can be recovered by either replaying the algorithm (by storing
the necessary randomization from the first run) or simply maintaining
the weights by persistent data structures and keeping a pointer to the
persistent copy of the weights from iteration $i^*$.

\end{document}

%% file: algos/held-karp-direct.tex
\begin{routine}{held-karp-1}{\defgraph}{\capacity}{\eps}
  $x \gets \zeroes^{\cuts}$, %
  $\weight \gets 1 / \capacity$, %
  $t \gets 0$,
  $\eta \gets \eps / \ln m$ \\
  while $t < 1$ \\
  \>
  \begin{math}
    C \gets \argmin_{C \in \cuts} \sumweight{C}
  \end{math}                    %
  \commentcode{compute the minimum cut}
  \\[.5ex]
  \>
  \begin{math}
    \varx \gets \frac{\ripof{\weight}{\capacity}}{\sumweight{C }}
    \cdot 1_C
  \end{math}
  \commentcode{%
    \begin{math}
      \varx = \argmaxof{ %
        \ripof{\ones}{\varvarx} %
        \where %
        \varvarx \in \nnreals^{\cuts} %
        \text{ and }
        \sum_{e \in E} \weight{e} \sum_{C \ni e} \varvarx{C} \leq
        \ripof{\weight}{\capacity}}
    \end{math}
  }\\
  \>
  \begin{math}
    \mincap \gets \min_{e \in C} \capacity{e}
  \end{math}\\[.5ex]
  \>
  \begin{math}
    \delta \gets              %
    \frac{\eps \mincap}{\eta}
  \end{math}
  \commentcode{such that
    \begin{math}
      \delta \sum_{C \ni e} \varx{C} %
      \leq                           %
      \frac{\eps}{\eta}
      \capacity{e}
    \end{math}
    for all $e \in \edges$ }
  \\
  \> $x \gets x + \delta y$ \commentcode{$x$ is maintained implicitly}
  \\
  \> for all $e \in C$ \\
  \> \>
  \begin{math}
    \weight{e}                                          %
    \gets                                               %
    \weight{e} \expof{\delta \eta / \capacity{e}} %
    =                                                              %
    \weight{e} \expof{\eps \mincap / \capacity{e}}
  \end{math}
  \commentcode{increase the weight of edge $e$}
  \\
  \> $t \gets t + \delta$ \\
  end while \\
  return $x$
\end{routine}

%%% Local Variables:
%%% mode: latex
%%% TeX-master: "../held-karp"
%%% End:

%% file: algos/held-karp-epochs.tex
\begin{routine}{held-karp-2}{\defgraph}{\capacity}{\eps}
  $x \gets \zeroes^{\cuts}$, $w \gets 1 / \capacity$, $t \gets 0$,
  $\eta \gets \eps / \ln m$
  \\
  $\lambda \gets $ size of minimum cut w/r/t edge weights
  $\weight$ \\
  while $t < 1$ \\
  \> while \= (a) $t < 1$ and \\
  \> \> (b) there is a cut $C \in \cuts$ s.t.\ $\sumweight{\capacity} \leq (1 + \eps) \lambda$ \\
  \retab %
  \> \> %
  \begin{math}
    \varx \gets \frac{\ripof{\weight}{\capacity}}{\sumweight{C }}
    \cdot 1_C
  \end{math}
  \commentcode{%
    \begin{math}
      \ripof{\ones}{\varx} \leq (1 + \eps) \maxof{ %
        \ripof{\ones}{\varvarx} %
        \where %
        \varvarx \in \nnreals^{\cuts} %
        \text{ and }
        \sum_{e \in E} \weight{e} \sum_{C \ni e} \varvarx{C} \leq
        \ripof{\weight}{\capacity}}
    \end{math}
  }\\
  \> \>
  \begin{math}
    \mincap \gets \min_{e \in C} \capacity{e}
  \end{math}\\[.5ex]
  \> \>
  \begin{math}
    \delta \gets              %
    \frac{\eps \mincap}{\eta}
  \end{math}
  \commentcode{such that
    \begin{math}
      \delta \sum_{C \ni e} \varx{C} %
      \leq                           %
      \frac{\eps}{\eta}
      \capacity{e}
    \end{math}
    for all $e \in \edges$
  }
  \\
  \> \> $x \gets x + \delta y$ \\
  \> \> for all $e \in C$ \\
  \> \> \>
  \begin{math}
    \weight{e} %
    \gets %
    \expof{\delta \eta / \capacity{e}} \cdot \weight{e} %
    = %
    \expof{\eps \mincap / \capacity{e}} \cdot \weight{e}
  \end{math} \\
  \> \> $t \gets t + \delta$\\
  \> end while \\
  \> $\lambda \gets (1 + \eps) \lambda$
  \commentcode{start new epoch} %
  \\
  end while \\
  return $x$
\end{routine}

%%% Local Variables:
%%% mode: latex
%%% TeX-master: "../held-karp"
%%% End:

%% file: algos/held-karp-trees.tex
\begin{routine}{held-karp-3}{\defgraph}{\capacity}{\eps}
  $x \gets \zeroes^{\cuts}$, %
  $\weight \gets 1 / \capacity$, %
  $t \gets 0$, $\eta \gets \eps / \ln m$ \\
  $\lambda \gets $ size of minimum cut w/r/t edge weights
  $\weight$ \\
  while $t < 1$ \\
  \> $\packing \gets$ $O(\logof{n/\eps})$ \= spanning trees capturing
  every
  $(1 + \eps)$-apx min cut \\
  \> \> w/ high probability \commentcode{by
    \reftheorem{karger-tree-packing}} \\
  \retab %
  \> for each tree $T \in \packing$ \\
  \> \> let $\cuts_T \subseteq \cuts$ be the cuts induced by
  removing $\leq 2$ edges from $T$ \\
  \> \> while \= (a) $t < 1$ and \\
  \> \> \> (b) there is a cut $C \in \cuts_T$ s.t.\ %
  $\sumweight{C} < (1 + \eps) \lambda$ \\
  \retab %
  \> \> \>
  \begin{math}
    y \gets \frac{\ripof{\weight}{\capacity}}{\sumweight{C}} \cdot C %
  \end{math}
  \\
  \> \> \>
  \begin{math}
    \mincap \gets \min_{e \in C} \capacity{e}
  \end{math} \\
  \> \> \>
  \begin{math}
    \delta \gets \frac{\eps \gamma}{\eta}
  \end{math}
  \\
  \> \> \> $x \gets x + \delta y$ \\
  \> \> \> for all $e \in C$ \\
  \>\>\>\>
  \begin{math}
    \weight{e} %
    \gets %
    \expof{\eps \gamma / \capacity{e}} \cdot \weight{e}
  \end{math} \\
  \> \> \> $t \gets t + \delta $\\
  \> \> end while \\
  \> end for \\
  \> $\lambda \gets (1 + \eps) \lambda$ %
  \commentcode{start new epoch}         %
  \\
  end while \\
  return $x$
\end{routine}

%%% Local Variables:
%%% mode: latex
%%% TeX-master: "../held-karp"
%%% End:

%% file: algos/held-karp-lazy-cuts.tex
\begin{routine}{held-karp-4}{\defgraph}{c}{\eps}
  $x \gets \zeroes^{\cuts}$, $\weight \gets \ones^{\edges}$,
  $t \gets 0$, $\eta \gets \eps / \ln m$\\
  $\lambda \gets$ size of minimum cut w/r/t edge weights
  $\weight$\\
  while $t < 1$\\
  \> $\packing \gets O(\logof{n / \eps})$ spanning trees capturing
  all $(1+\eps)$-apx min cuts \\
  \> \> w/r/t weights $w$ w/ high probability %
  \commentcode{by \reftheorem{karger-tree-packing}} \\
  \retab %
  \> for each tree $T \in \packing$\\
  \> \> $\refroutine{lic} \gets \refroutine{lazy-inc-cuts.init}{\graph}{\capacity}{T}$\\
  \> \> let $\cuts_T \subseteq \cuts$ be the cuts induced by
  removing $\leq 2$ edges from $T$ \\
  \> \> while \= (a) $t < 1$ and until \\
  \> \> \> (b) \= we have certified there are no cuts
  $C \in \cuts_T$
  \\
  \> \> \> \> %
  s.t.\
  \begin{math}
    \sumweight{C} < (1 + \eps) \lambda
  \end{math}
  \\
  \retab %
  \> \> \> find $C \in \cuts_T$ w/
  \begin{math}
    \sumweight{e} %
    < %
    (1 + O(\eps)) \lambda
  \end{math} \\
  \> \> \> %
  \begin{math}
    y \gets %
    \prac{\ripof{\weight}{\capacity}}%
    {\sumweight{C}} %
    \cdot %
    C
  \end{math}\\
  \> \> \> %
  \begin{math}
    \mincap \gets \min_{e \in C} \capacity{e}
  \end{math}\\[.5ex]
  \> \> \> %
  \begin{math}
    \delta \gets \frac{\eps \mincap}{\eta}
  \end{math}\\
  \> \> \>
  \begin{math}
    x \gets x + \delta y
  \end{math}\\
  \> \> \>
  \begin{math}
    t \gets t + \delta
  \end{math} \\
  \> \> \>
  \begin{math}
    \Delta \gets \refroutine{lic.inc-cut}{C}
  \end{math} \\
  \> \> \> for $(e,\delta) \in \Delta$ \\
  \> \> \> \> $\weight{e} \gets \weight{e} + \delta$ \\
  \> \> \> end for \\
  \> \> end while \\
  \> \> $\Delta \gets \refroutine{lic.flush}{}$\\
  \> \> for $(e,\delta) \in \Delta$\\
  \> \> \> $\weight{e} \gets \weight{e} + \delta$ \\
  \> \> end for \\
  \> end for \\
  end while \\
  return $x$
\end{routine}

%%% Local Variables:
%%% mode: latex
%%% TeX-master: "../held-karp"
%%% End: